\documentclass[
aps,
prab,
reprint,
amsmath,
amssymb,
superscriptaddress,
longbibliography
]{revtex4-1}
\usepackage{xcolor}
\usepackage{graphicx}

\begin{document}

\title{High-power attosecond X-ray free-electron lasers: physics and design strategy}

\author{Chenzhi Xu}
\affiliation{Shanghai Institute of Applied Physics, Chinese Academy of Sciences, Shanghai 201800, China}
\affiliation{University of Chinese Academy of Sciences, Beijing 100049, China}
\affiliation{European XFEL, 22869 Schenefeld, Germany}

\author{Jiawei Yan}
\email{jiawei.yan@desy.de}
\affiliation{Deutsches Elektronen-Synchrotron DESY, 22603 Hamburg, Germany}

\author{Ye Chen}

\affiliation{Deutsches Elektronen-Synchrotron DESY, 22603 Hamburg, Germany}

\author{Winfried Decking}

\affiliation{Deutsches Elektronen-Synchrotron DESY, 22603 Hamburg, Germany}

\author{Marc Guetg}

\affiliation{Deutsches Elektronen-Synchrotron DESY, 22603 Hamburg, Germany}

\author{Tianyun Long}

\affiliation{Deutsches Elektronen-Synchrotron DESY, 22603 Hamburg, Germany}

\author{Bingyang Yan}

\affiliation{Shanghai Institute of Applied Physics, Chinese Academy of Sciences, Shanghai 201800, China}
\affiliation{University of Chinese Academy of Sciences, Beijing 100049, China}

\author{Haixiao Deng}
\email{denghx@sari.ac.cn}
\affiliation{Shanghai Advanced Research Institute, Chinese Academy of Sciences, Shanghai 201210, China}

\begin{abstract}

Attosecond pulses from X-ray free-electron lasers (XFELs) have opened new opportunities for probing ultrafast electronic dynamics on the \AA ngstrom--attosecond spatiotemporal scale. Most attosecond XFEL concepts rely on generating an ultrashort high-current spike through either external laser modulation or accelerator-based beam manipulation. Despite their different implementations, these approaches share the same essential physics, namely that the XFEL amplification is confined to a short effective lasing window within the electron beam. However, existing studies are often scheme-specific and do not yet provide a unified quantitative picture of how fundamental electron-beam properties constrain high-power attosecond performance. In this work, we investigate the general physics and scheme-independent requirements for generating high-power attosecond X-ray pulses from a short current spike. From the perspective of post-saturation superradiant evolution, we show that the electron-beam length governs both the attainable peak power and the pulse duration. We further examine the distinct roles of slice energy spread, slice emittance, energy chirp, and transverse beam tilt. Our results reveal the trade-off between peak power, pulse shortening, and single-spike probability, and provide facility-independent guidelines for optimizing electron-beam phase-space manipulation toward terawatt-class attosecond XFEL operation.

\end{abstract}

\maketitle

\section{Introduction}

The advent of attosecond X-ray pulses has opened new opportunities for probing ultrafast electronic dynamics in atoms, molecules, and condensed matter systems, enabling access to fundamental processes such as photoionization~\cite{driver2024attosecond} and charge migration as well as electronic coherence~\cite{goulielmakis2010realtime,kraus2015chargemigration,pertot2017xrayabsorption}. At present, attosecond X-ray sources are primarily realized via high-harmonic generation (HHG) and X-ray free-electron lasers (XFELs). HHG has demonstrated remarkable progress in generating attosecond pulses with excellent temporal coherence. However, the attainable photon energies are typically confined to the extreme-ultraviolet and soft X-ray regimes, making extension toward hard X rays challenging~\cite{paul2001science,hentschel2001nature,mashiko2008prl,krausz2009rmp,vincenti2012prl}.

XFELs extend attosecond science into a regime of much higher photon energy and pulse energy~\cite{yan2024terawattfel,t8qq-h31p,franz2024terawatt,serkez_fel2022_tuai2,huang2021innovation,Berrah:25}. During the last few years, attosecond XFEL pulses have been demonstrated in the soft- and hard X-ray regimes. Extending attosecond techniques into the hard X-ray domain is particularly significant~\cite{Huang2017PRL,yan2024terawattfel,t8qq-h31p,inoue2025experimental}, as it enables element-specific, core-level–resolved investigations of ultrafast dynamics in complex materials. It also motivates applications such as attosecond XFEL pump-probe spectroscopy and ultrafast structural imaging~\cite{zhu2024attosecondvision}.

In an XFEL, a relativistic electron beam interacts with a copropagating radiation field, initiated either from intrinsic shot noise or from an external seed, and the radiation is amplified along a long undulator to produce intense X-ray pulses~\cite{RevModPhys.88.015006}. Because the radiation pulse duration is determined by the effective lasing window within the electron bunch, most attosecond XFEL schemes aim to localize the amplification to an ultrashort fraction of the bunch, typically realized as a high-current spike. Other approaches, including FEL mode locking~\cite{Thompson2008PRL,wn8d-l7sh}, coherent synthesis of harmonics~\cite{Maroju2020Nature}, and single-cycle pulse generation using a tailored undulator~\cite{RebernikRibic:20}, can also produce attosecond pulses. The approaches considered here, which rely on a short effective lasing window, can be broadly categorized into laser-assisted methods and accelerator-based beam-manipulation methods.

On the laser-assisted side, the central idea is to use an external optical field to create a strong local modulation in the electron beam and then convert it into a short current spike. The basic modulation-and-dispersion picture first appeared in seeded FEL schemes such as high-gain harmonic generation (HGHG)~\cite{PhysRevA.44.5178}. The enhanced self-amplified spontaneous emission (ESASE) scheme later extended that idea to much stronger local compression by using an intense few-cycle or single-cycle laser~\cite{zholents2005method}. Related variants include self-modulation~\cite{Duris2020NatPhot}, two-laser modulation~\cite{ZholentsPenn2005,Ding2009ESASE}, optical-laser bunch compression~\cite{Duris2020NatPhot}, and sub-cycle laser modulation~\cite{xiao2025isolated}. 

Attosecond pulse generation can also be realized without external lasers, by using the accelerator to shape the longitudinal phase space of the electron beam and create a short current spike. Early studies emphasized very low bunch charge and aggressive compression as a route toward single-spike self-amplified spontaneous emission (SASE) operation~\cite{Rosenzweig2008NIMA,Reiche2008NIMA,t8qq-h31p}. That line of work was followed by experimental demonstrations based on low charge and nonlinear compression in soft- and hard X-ray FELs~\cite{Huang2017PRL,malyzhenkov2020attosecond,Prat2023APLP}. More recently, terawatt-level hard X-ray attosecond operation has been proposed based on higher-charge beams combined with self-chirping, with particular emphasis on strong pre-undulator compression in an arc or switchyard~\cite{yan2024terawattfel,yan2025attoshine,mnwd-ypfs}. Other approaches, including cathode shaping and laser-heater shaping, still follow the same basic principle of creating a sufficiently short current spike through energy modulation and electron-beam compression~\cite{Zhang2020NJP,Li2024APL}. In addition, Ref.~\cite{10.1063/5.0050693} explored the feasibility of terawatt-scale attosecond X-ray pulse generation in a plasma-accelerator-based scheme through the formation of a high-current spike.

These studies have clearly demonstrated that isolated attosecond X-ray pulses can be generated by confining FEL amplification to an ultrashort high-current spike. Different methods can lead to different output characteristics. Some achieve a high single-spike probability but relatively low peak power, whereas others produce attosecond pulses with high peak power. Previous numerical and experimental studies have mainly examined how machine settings and beam-manipulation parameters affect current-spike formation and XFEL performance in specific configurations. For example, Refs.~\cite{Shim2018,Shim2020} analyzed terawatt-level attosecond pulse generation in ESASE-based schemes. The FEL dynamics of a short chirped electron beam were analyzed in Ref.~\cite{PhysRevAccelBeams.21.110702}, but only in the linear regime before saturation. Taken together, existing studies do not yet provide a unified physical picture linking the underlying FEL dynamics to the resulting attosecond pulse characteristics.

In this work, we focus on the physics of the ultrashort current spike itself. We investigate how the electron-bunch length governs both the attosecond pulse duration and the attainable peak power from the perspective of post-saturation superradiance. We further examine the distinct effects of other electron-beam properties, including the slice energy spread, transverse emittance, energy chirp, and transverse beam tilt. Our goal is to provide a clear physical picture and more general design principles for future high-power attosecond XFELs.

\section{Post-saturation superradiant evolution in a single-stage SASE FEL }

Many attosecond XFEL concepts rely on a localized high-current spike to generate an ultrashort radiation pulse. In such schemes, the radiation typically starts from shot noise and evolves in the SASE regime. To clarify the essential role of slippage and the physical origin of post-saturation superradiant evolution, we first recall the one-dimensional time-dependent FEL equations~\cite{bonifacio1990superradiant,Bonifacio1994spectrum}:
\begin{align}
\frac{\partial \theta_j}{\partial \bar{z}} &= p_j, \\
\frac{\partial p_j}{\partial \bar{z}} &= -\left( A e^{i\theta_j} + \mathrm{c.c.} \right), \\
\frac{\partial A}{\partial \bar{z}}+\frac{\partial A}{\partial \bar{z}_1} &= \left\langle e^{-i\theta_j} \right\rangle ,
\end{align}
Here $z$ is the longitudinal distance along the undulator and $t$ is the time in the laboratory frame. The scaled undulator distance is defined as $\bar{z}=z/l_g$, while the longitudinal coordinate within the bunch is defined as $\bar{z}_1=-c(t-z/v_{\parallel})/l_c$, where $v_{\parallel}$ is the longitudinal electron velocity and $c$ is the speed of light. The quantities $l_g=\lambda_w/(4\pi\rho)$ and $l_c=\lambda_s/(4\pi\rho)$ are the gain length and the cooperation length, respectively. Here $\lambda_w$ denotes the undulator period. The radiation wavelength satisfies the resonance condition $\lambda_s=\lambda_w(1+K^2/2)/(2\gamma_0^2)$, where $K$ is the undulator parameter and $\gamma_0$ is the reference Lorentz factor. The parameter $\rho$ is the Pierce parameter~\cite{BONIFACIO1984373}. The field variable $A$ is the normalized radiation amplitude, $\theta_j=(k_w+k_s)z-ck_st_j$ is the ponderomotive phase of the $j$th electron, and $p_j$ is its normalized energy deviation. Here $k_w$ and $k_s$ are the undulator and radiation wavenumbers, respectively. These equations are introduced here to illustrate the underlying one-dimensional physics. The quantitative results discussed below are obtained from three-dimensional time-dependent simulations.

\begin{figure*}[t]
\centering
\includegraphics[width=0.9\linewidth]{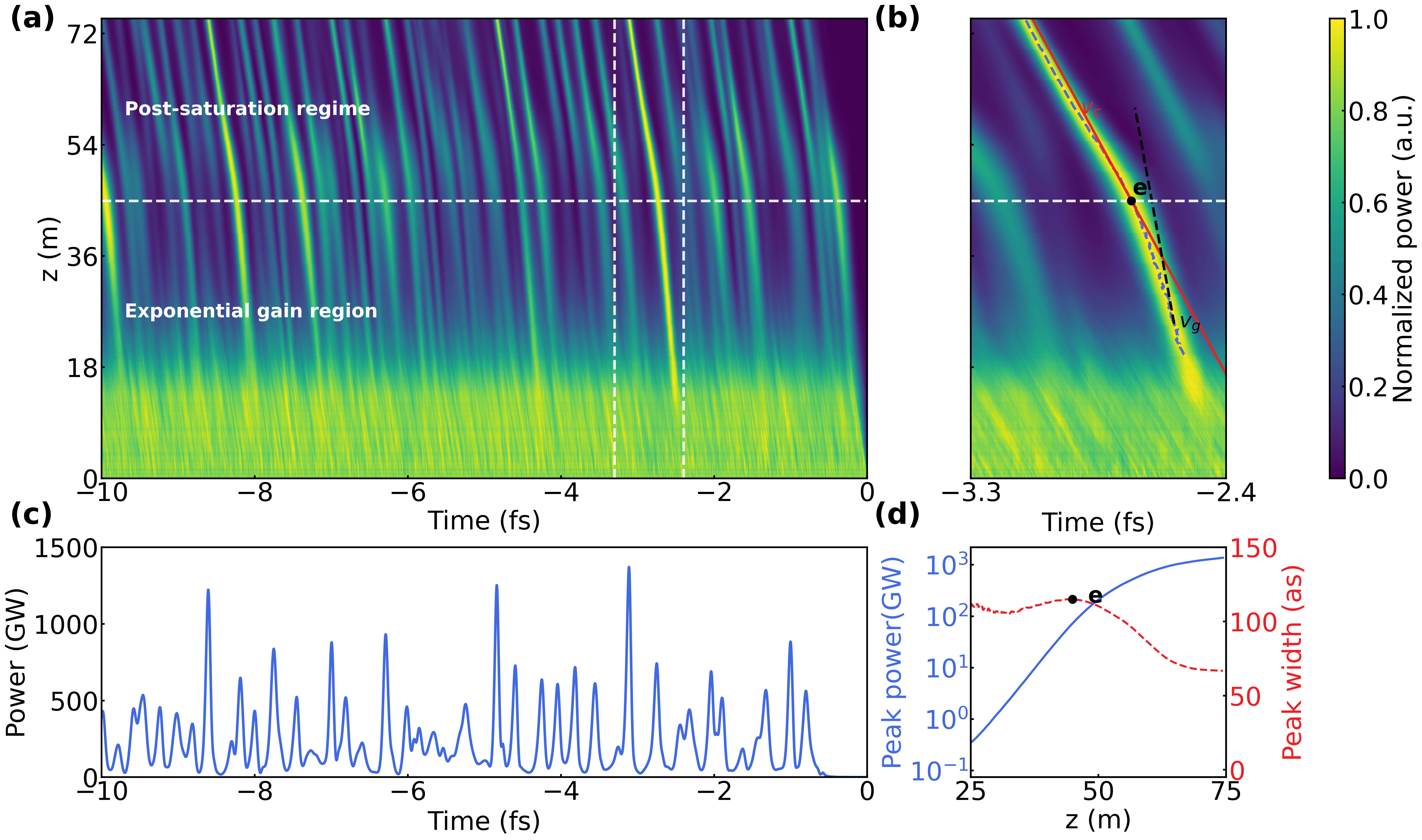}
\caption{(a) Two-dimensional evolution of the radiation power along the undulator. At each undulator position $z$, the power profile is normalized to its local maximum. (b) A representative spike extracted for further analysis. The red line indicates the speed of light, the black dashed line shows the group velocity $v_g=c(1-2\lambda_s/3\lambda_w)$, and the blue dashed line marks the spike peak. (c) Temporal radiation-power profiles at the end of the fifteenth undulator segment. (d) Evolution of the spike peak power and peak width along the undulator for the spike shown in panel (b). The position labeled $e$ in panel (d), corresponding to $z\approx 45~\mathrm{m}$ where the spike width begins to decrease, is indicated by the white transverse line in panels (a) and (b). The bunch head is to the left.}
\label{Fig:part0_long_bunch}
\end{figure*}

The term $\partial A/\partial \bar{z}_1$ represents radiation slippage, whereby the pulse slips forward by one radiation wavelength per undulator period. If this term is neglected, the equations reduce to the steady-state one-dimensional FEL equations, for which saturation is followed by the familiar post-saturation power oscillations~\cite{BONIFACIO1984373,PhysRevSTAB.10.034801,PhysRevAccelBeams.23.120703,10.1063/5.0166336}. Once slippage is retained, however, the pulse evolution becomes explicitly time dependent~\cite{bonifacio1990superradiant}. In the exponential-gain regime, the radiation still amplifies exponentially while propagating with the group velocity $v_g=c(1-2\lambda_s/3\lambda_w)$, and the temporal spike width increases approximately as $\sqrt{z}$. 

The key difference appears after saturation, when slippage allows a strong spike to undergo post-saturation superradiant evolution~\cite{Bonifacio1994spectrum}. In this regime, the radiation field $A$ in Eq.~(3) becomes large enough to drive rapid phase-space rotation of the electrons through the ponderomotive force in Eq.~(2), leading to efficient energy transfer from the electron beam to the leading edge of the radiation spike, while the trailing edge returns energy to the electrons. Combined with the slippage term $\partial A/\partial \bar{z}_1$, this phase-space dynamics concentrates the radiation energy within an ever-narrowing temporal window, producing the simultaneous peak-power growth and spike-width compression characteristic of post-saturation superradiance~\cite{bonifacio1990superradiant,Bonifacio1994spectrum, PhysRevLett.98.034802,giannessi2005nonlinear,yang2020postsaturation,Mirian2021,PONGCHALEE2024107673,yan2025attoshine}. In the superradiant regime, the group velocity also departs from the exponential-gain value and approaches the speed of light~\cite{Bonifacio1994spectrum,yang2020postsaturation}. Three-dimensional superradiant dynamics were further analyzed in Ref.~\cite{robles2024three}. Ref.~\cite{yan2025attoshine} suggests that such post-saturation superradiant evolution in single-stage SASE operation may be central to the generation of terawatt attosecond XFEL pulses.

To illustrate the post-saturation superradiant behavior, we perform a three-dimensional time-dependent SASE simulation. We deliberately choose a $10~\mathrm{fs}$ flat-top bunch so that the full evolution from exponential gain to post-saturation superradiance can be observed in a single run, rather than to represent an optimized attosecond operating point. The electron beam has an energy of $14~\mathrm{GeV}$, a peak current of $20~\mathrm{kA}$, a normalized emittance of $0.5~\mathrm{\mu m}$, an average $\beta$ function of $32~\mathrm{m}$, and a slice energy spread of $10~\mathrm{MeV}$. The undulator period is $40~\mathrm{mm}$, each segment is $5~\mathrm{m}$ long, and the photon energy is $8700~\mathrm{eV}$. The gaps between undulator segments are included in the simulations, but are not shown in the figure in order to present the physical evolution more clearly. A total of 15 segments are tracked using GENESIS~1.3 (version~4)~\cite{reiche1999genesis,reiche2022genesis4,reiche2024genesis}. For this configuration the Pierce parameter is $\rho=1.07\times10^{-3}$, and the saturation power estimated from the Ming Xie fitting formulas~\cite{xie1995design} is approximately $180~\mathrm{GW}$. Throughout this work, no post-saturation taper is applied, and the only tapering considered is the chirp-compensating taper introduced later for chirped beams. Dedicated post-saturation taper optimization, which would further enhance peak power, is beyond the scope of this work.

As shown in Fig.~\ref{Fig:part0_long_bunch}, after saturation the SASE spike continues to grow, with its peak power rising well above the saturation power. To examine this process in more detail, we follow a representative spike, as shown in Fig.~\ref{Fig:part0_long_bunch}(b) and~(d). The peak width, defined as the full width at half maximum (FWHM) of the local power spike, begins to decrease at point~$e$, corresponding to $z\approx 45~\mathrm{m}$. Beyond this point, the peak width narrows from $115~\mathrm{as}$ to $67~\mathrm{as}$, while the peak power increases from $72~\mathrm{GW}$ to $1.3~\mathrm{TW}$, far above the saturation power estimated from the Ming Xie fitting formulas. At the same time, the radiation group velocity approaches the speed of light. Taken together, these signatures indicate the onset of post-saturation superradiant evolution. They also highlight the importance of this regime for attosecond pulse generation, since it enables both substantially higher peak power and a narrower dominant spike.

Slippage is at the heart of SASE~\cite{Bonifacio1994spectrum,BONIFACIO2005645}, and is even more critical for attosecond XFEL pulse generation. Because of radiation slippage, entering the post-saturation regime requires that the radiation pulse remain inside the electron beam long enough for continued amplification. If the bunch is too short, the radiation pulse slips out of the beam before entering the post-saturation regime. This makes the bunch length a key parameter for attosecond XFEL generation. We therefore carry out a systematic study of the dependence of attosecond XFEL performance on the electron-bunch length. Time-dependent simulations are performed for 50 shots at each bunch length using a total of 15 undulator segments. Unless otherwise stated, the electron-beam parameters are the same as those given above, and the undulator period is kept unchanged.

\begin{figure}[!htb]
\centering
\includegraphics[width=0.9\linewidth]{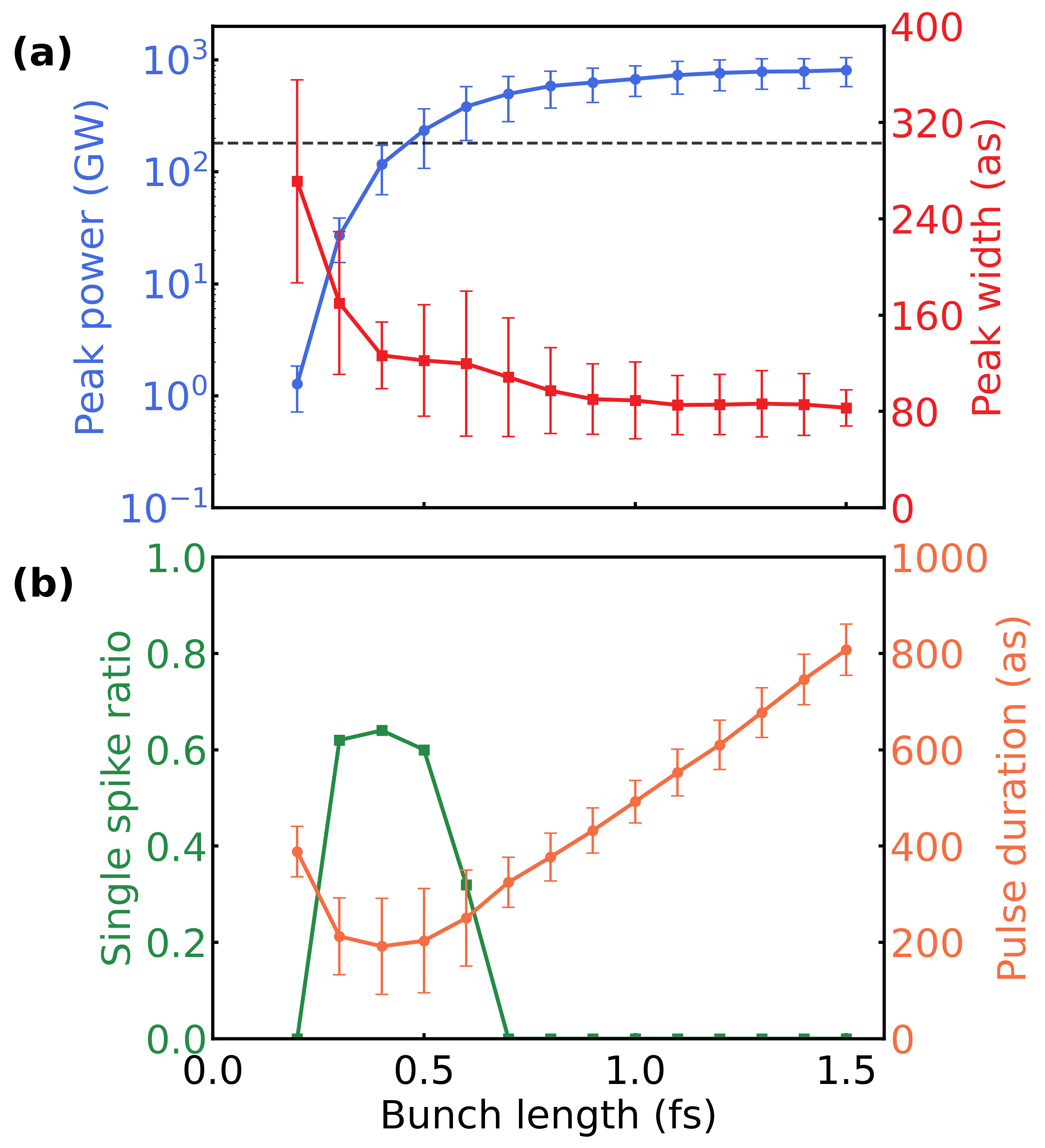}
\caption{Dependence of FEL pulse properties on the electron-bunch length. (a) Peak power and peak width. The horizontal black line indicates the saturation power calculated from the Ming Xie fitting formulas~\cite{xie1995design}. (b) Pulse duration and single-spike ratio. Error bars denote the shot-to-shot standard deviation obtained from 50 simulations at each bunch length.}
\label{Fig:part1_bunch_length_mean_peak_power}
\end{figure}

As shown in Fig.~\ref{Fig:part1_bunch_length_mean_peak_power}(a), the mean peak power increases with bunch length. Around 500~as, the mean peak power exceeds the calculated saturation power. Below this point, the peak power increases nearly exponentially with bunch length. Beyond it, the peak power continues to grow approximately linearly with bunch length. At a bunch length of $1~\mathrm{fs}$, the mean peak power reaches $676~\mathrm{GW}$, while the 90th percentile of the peak-power distribution is $0.94~\mathrm{TW}$. The peak width shows the opposite trend. As the pulse enters the post-saturation regime, superradiant evolution compresses the main spike, reducing the peak width from $122~\mathrm{as}$ at a bunch length of $500~\mathrm{as}$ to $89~\mathrm{as}$ at $1~\mathrm{fs}$.

For applications that require a clean single attosecond pulse, we further examine the single-spike ratio. A shot is classified as single-spike if a standard peak-finding algorithm identifies only one peak in its temporal power profile. Here, peaks are defined as local maxima with heights exceeding 1/20 of the global power maximum and prominences exceeding 1/30 of the global power maximum. The single-spike ratio is then defined as the fraction of such shots in the statistical ensemble. For bunch lengths around 300--500~as, the single-spike ratio remains relatively high, exceeding 60\%. As the bunch length increases further, the single-spike ratio decreases, indicating a growing tendency toward multi-spike structures. We also analyze the pulse duration, which characterizes the overall width of the radiation pulse. For a single-spike pulse, it is taken directly as the FWHM. For a multi-spike pulse, we construct an equivalent Gaussian envelope from the signal-weighted first and second moments of the temporal profile and use the FWHM of that envelope. The pulse duration reaches its minimum for bunch lengths around 300--500~as. Although the dominant spike becomes narrower as the bunch length increases, the overall pulse duration becomes longer for bunch lengths above 500~as because the number of spikes increases. In addition, the shot-to-shot jitter of the pulse duration becomes smaller in this longer-bunch regime.

\begin{figure}[!htb]
\centering
\includegraphics[width=0.9\linewidth]{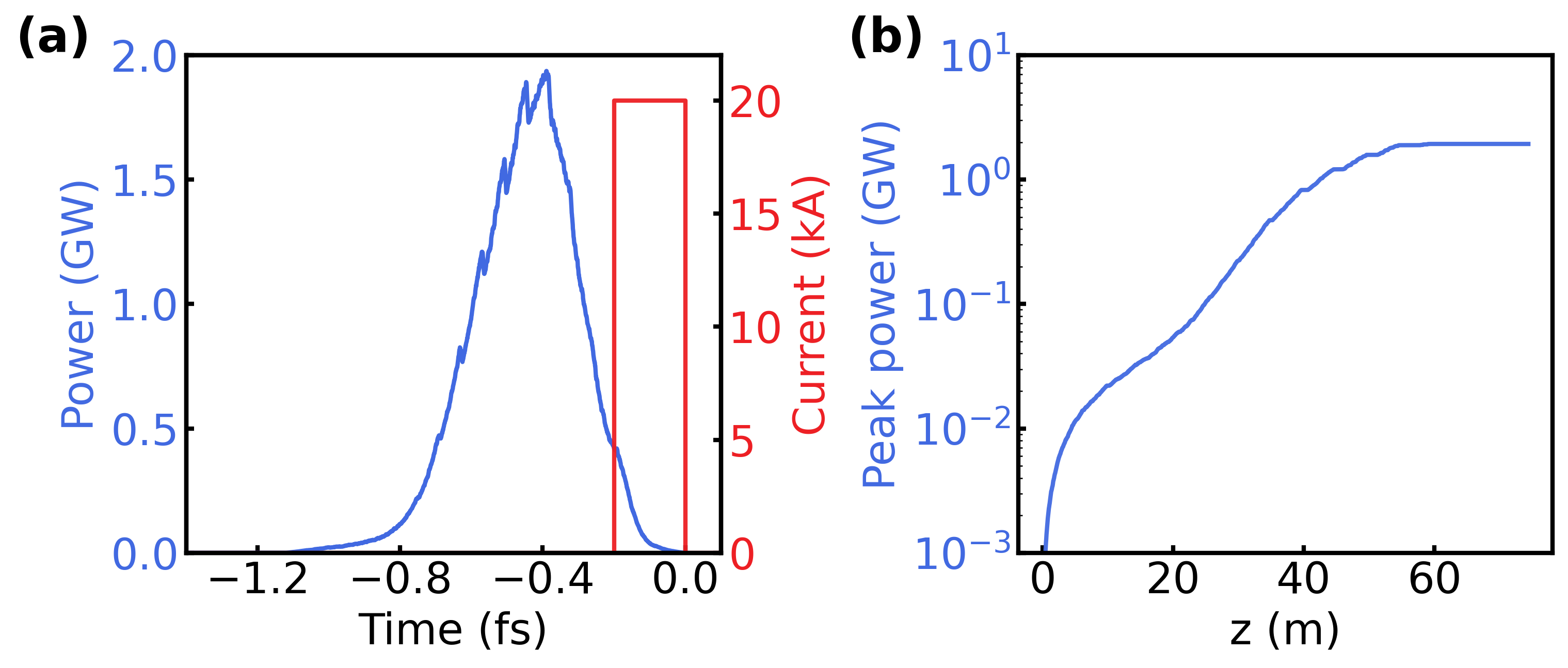}
\caption{(a) Temporal profiles of the FEL radiation power (blue) and the electron-beam current (red) at the end of the fifteenth undulator segment for a bunch length of 200~as. (b) Evolution of the peak power along the undulator.}
\label{Fig:part1_short_electron_beam_200}
\end{figure}

For bunch lengths below $300~\mathrm{as}$, further shortening of the bunch does not improve the single-spike ratio or reduce the pulse duration. In this regime, the lasing region is too short for the radiation field to remain overlapped with the electron beam long enough to achieve substantial amplification. Once the radiation slips out of the bunch, the amplification is interrupted before a dominant spike can fully develop. The microbunching imprinted on the beam can still emit radiation and seed further growth, but the radiation repeatedly outruns the short lasing region, leading to weak and poorly isolated spike structures. A representative $200~\mathrm{as}$ case is shown in Fig.~\ref{Fig:part1_short_electron_beam_200}, where the pulse contains many small spikes. At the end of the undulator, the peak power is only $2~\mathrm{GW}$ and the pulse duration reaches $349~\mathrm{as}$.

\begin{figure}[!htb]
\centering
\includegraphics[width=0.9\linewidth]{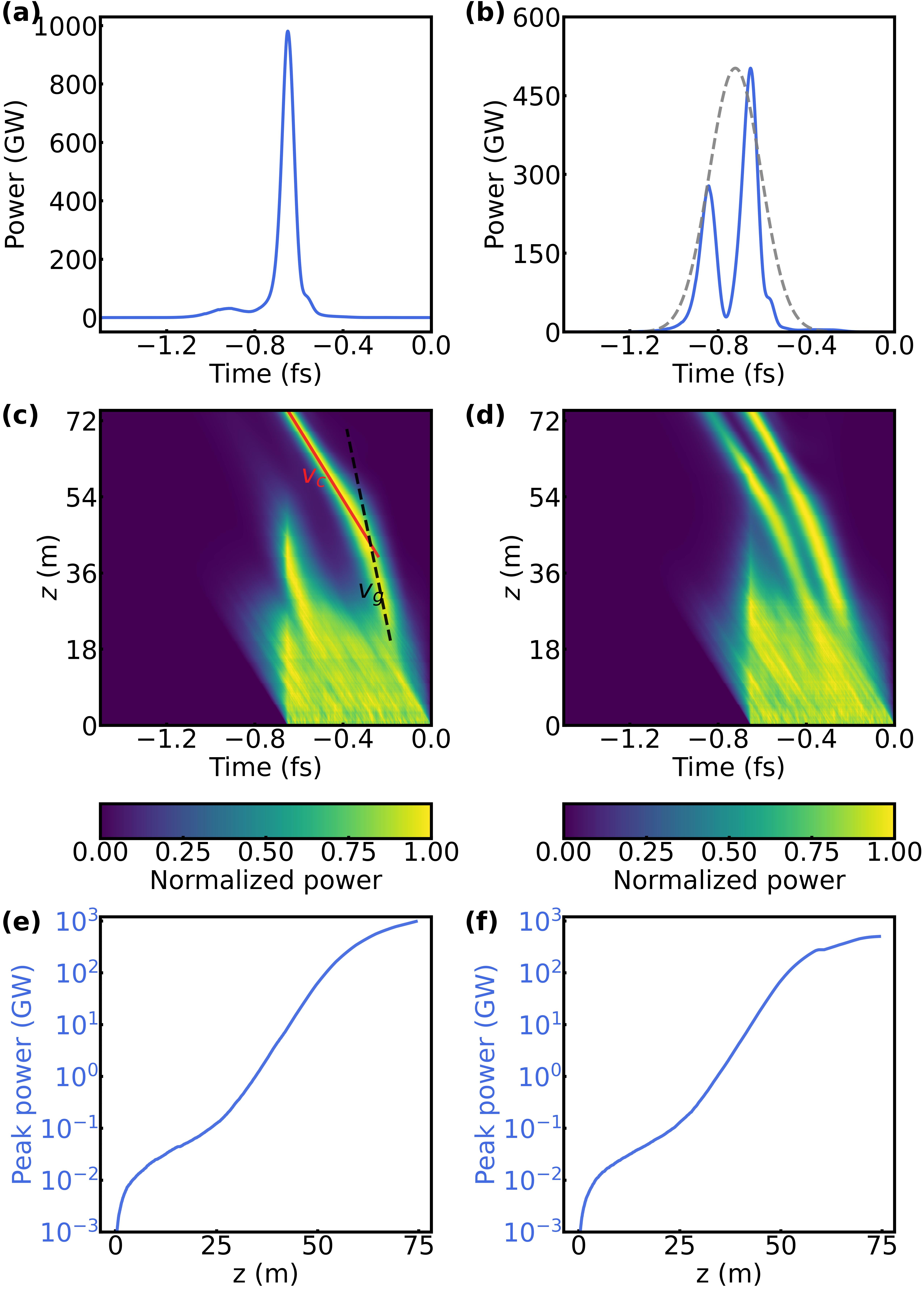}
\caption{Representative shots for an electron-bunch length of $650~\mathrm{as}$. Left column: a shot with a dominant spike. Right column: a shot with multiple spikes. (a) and (b) Temporal profiles of the FEL radiation power (blue) and the corresponding envelope (gray) at the end of the fifteenth undulator segment. (c) and (d) Two-dimensional evolution of the radiation power along the undulator. At each undulator position $z$, the power profile is normalized to its local maximum. (e) and (f) Evolution of the peak power along the undulator.}
\label{Fig:part1_long_electron_beam_650as}
\end{figure}

These results indicate that the electron-bunch length should be chosen according to the FEL dynamics one aims to exploit. Physically, the bunch length is a primary parameter controlling the effective lasing window over which radiation remains coupled to the electrons through slippage, and therefore the balance between single-spike formation and access to the post-saturation superradiant regime. For bunch lengths of about $300$--$500~\mathrm{as}$, the dominant spike can be amplified to a substantial level, while the lasing window remains short enough to suppress pronounced multi-spike development. This regime is therefore favorable for producing a reasonably single spike. For longer bunches, the radiation remains coupled to the electron beam over a larger slippage distance, allowing the dominant spike to propagate further into the post-saturation regime. In this regime, continued energy extraction from the electrons drives superradiant evolution of the leading spike, resulting in increasing peak power and decreasing spike width. From this perspective, bunch lengths above $500~\mathrm{as}$, and even up to $\sim1~\mathrm{fs}$, are advantageous when the primary objective is to maximize photon number or peak power, as required in nonlinear X-ray experiments or other photon-hungry applications. The trade-off is that the longer lasing window also allows secondary spikes to develop concurrently with the dominant spike, and trailing radiation structures to form as the dominant spike slips forward through the modulated electron bunch, both of which reduce the probability of obtaining a clean single-spike pulse. The analysis of this physical process motivated us to investigate high-power hard X-ray attosecond pulse generation using high-charge electron beams~\cite{yan2024terawattfel,yan2025attoshine}. The self-chirping operation employed there achieves a high peak current while maintaining a suitable electron-bunch length, thereby providing favorable conditions for post-saturation superradiant evolution.

A particularly interesting intermediate regime occurs around $650~\mathrm{as}$, where the bunch is long enough for the dominant spike to already benefit from post-saturation superradiant compression, yet still short enough that single-spike shots remain possible. As shown in Fig.~\ref{Fig:part1_long_electron_beam_650as}(a), the main spike stays overlapped with the electron beam over a sufficiently long distance to undergo further post-saturation growth, resulting in both a higher peak power and a narrower temporal width. In a representative shot, the peak power approaches $1~\mathrm{TW}$ while the pulse duration decreases to about $66~\mathrm{as}$. At the same time, the longer interaction window also permits the appearance of multi-spike shots, as shown in Fig.~\ref{Fig:part1_long_electron_beam_650as}(b). The simulations presented in the main text use a flat-top longitudinal current profile. Additional simulations using a Gaussian current profile are presented in Appendix~\ref{sec:gaussian_profile}. 

\begin{figure*}[t]
\centering
\includegraphics[width=0.9\linewidth]{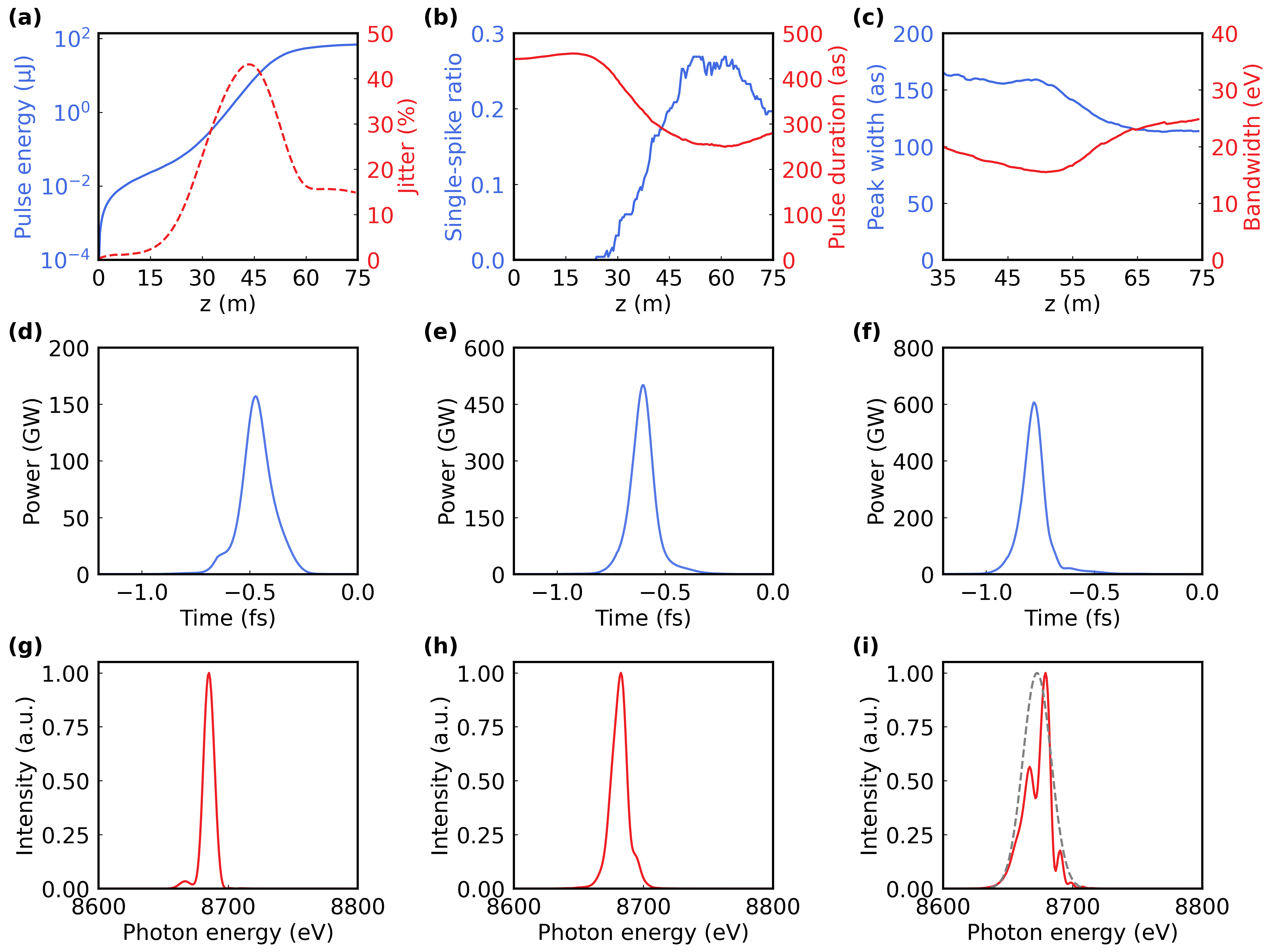}
\caption{Statistical analysis for the $650~\mathrm{as}$ bunch-length case. (a) Mean pulse energy and relative pulse-energy jitter as functions of undulator distance. (b) Temporal single-spike ratio and mean pulse duration as functions of undulator distance. (c) Mean temporal peak width and mean spectral bandwidth as functions of undulator distance, evaluated from the point where the dominant spike becomes clearly identifiable. (d)--(f) Temporal power profiles of a temporal single-spike shot at $z=50$, $60$, and $75~\mathrm{m}$, respectively. (g)--(i) Corresponding spectra at $z=50$, $60$, and $75~\mathrm{m}$, respectively. The gray dashed lines indicate the spectral envelopes.}
\label{Fig:part1_650as_jitter_analysis}
\end{figure*}

The shot-to-shot behavior of the $650~\mathrm{as}$ case is worth examining in more detail. To this end, we perform 200 shot-noise simulations with GENESIS~1.3 (version~4). At the undulator exit, the mean peak power is $449~\mathrm{GW}$ and the mean pulse duration is $278~\mathrm{as}$. The 90th percentile of the peak-power distribution is $0.68~\mathrm{TW}$. For the 37 shots classified as single-spike at the undulator exit, the mean peak power and mean pulse duration are $557~\mathrm{GW}$ and $100~\mathrm{as}$, respectively. For these single-spike shots, the dominant spike contains, on average, $93\%$ of the total pulse energy, evaluated using the nearest crossings at $5\%$ of the peak power as the integration boundaries. As shown in Fig.~\ref{Fig:part1_650as_jitter_analysis}(a), the mean pulse energy continues to increase along the undulator, whereas the relative pulse-energy jitter first increases, reaches a maximum around $z=45~\mathrm{m}$, and then decreases toward the end of the undulator. The reduction in jitter continues even after the pulse enters the post-saturation regime. Figure~\ref{Fig:part1_650as_jitter_analysis}(b) shows the evolution of the temporal single-spike ratio and mean pulse duration along the undulator. As the undulator length increases, the single-spike ratio first rises and reaches a maximum of about $24$--$25\%$ in the range $z=50$--$60~\mathrm{m}$. Over the same range, the mean pulse duration decreases and reaches its minimum near the position of the maximum single-spike ratio. With further propagation, the single-spike ratio decreases and the mean pulse duration begins to increase. The spectral bandwidth initially decreases and then broadens after approximately $z=50~\mathrm{m}$, as shown in Fig.~\ref{Fig:part1_650as_jitter_analysis}(c). For a single-spike spectrum, the bandwidth is defined as the FWHM of the spectral spike itself, whereas for a multi-spike spectrum it is taken as the FWHM of an equivalent Gaussian spectral envelope.

Once the pulse enters the post-saturation regime, the dominant spike continues to grow and propagate, while its width decreases owing to superradiant evolution. Meanwhile, continued radiation from the beam regions behind the dominant spike, together with the growth of secondary structures, produces a trailing temporal profile. As a result, the single-spike ratio subsequently decreases to 18.5\% at the end of the fifteenth undulator segment. At the same time, Fig.~\ref{Fig:part1_650as_jitter_analysis}(d)--(i) shows the temporal and spectral evolution of a temporal single-spike shot. From $z=50$ to $75~\mathrm{m}$, the main spike continues to compress and increase in peak power, while weak trailing structures gradually develop. At $z=60~\mathrm{m}$, both the temporal profile and the corresponding spectrum are classified as single-spike. At $z=75~\mathrm{m}$, the weak trailing structures do not form an additional temporal peak satisfying the peak-detection criteria, and the temporal profile remains classified as single-spike, whereas the corresponding spectrum develops three distinct peaks. At present, single-shot temporal diagnostics for hard X-ray attosecond pulses remain challenging~\cite{inoue2025experimental,sun2025ultrastablehardxrayattosecond}. In experiments, pulse duration is often inferred from the bandwidth of a single-spike spectrum, and machine tuning is commonly guided by the appearance of such spectra. The present results suggest that this strategy may miss the highest-power operating point, because post-saturation enhancement can already occur after the spectrum has ceased to be cleanly single-spike. Single-shot temporal diagnostics therefore remain necessary when optimizing for the highest-power attosecond pulses.

\begin{figure}[!htb]
\centering
\includegraphics[width=0.9\linewidth]{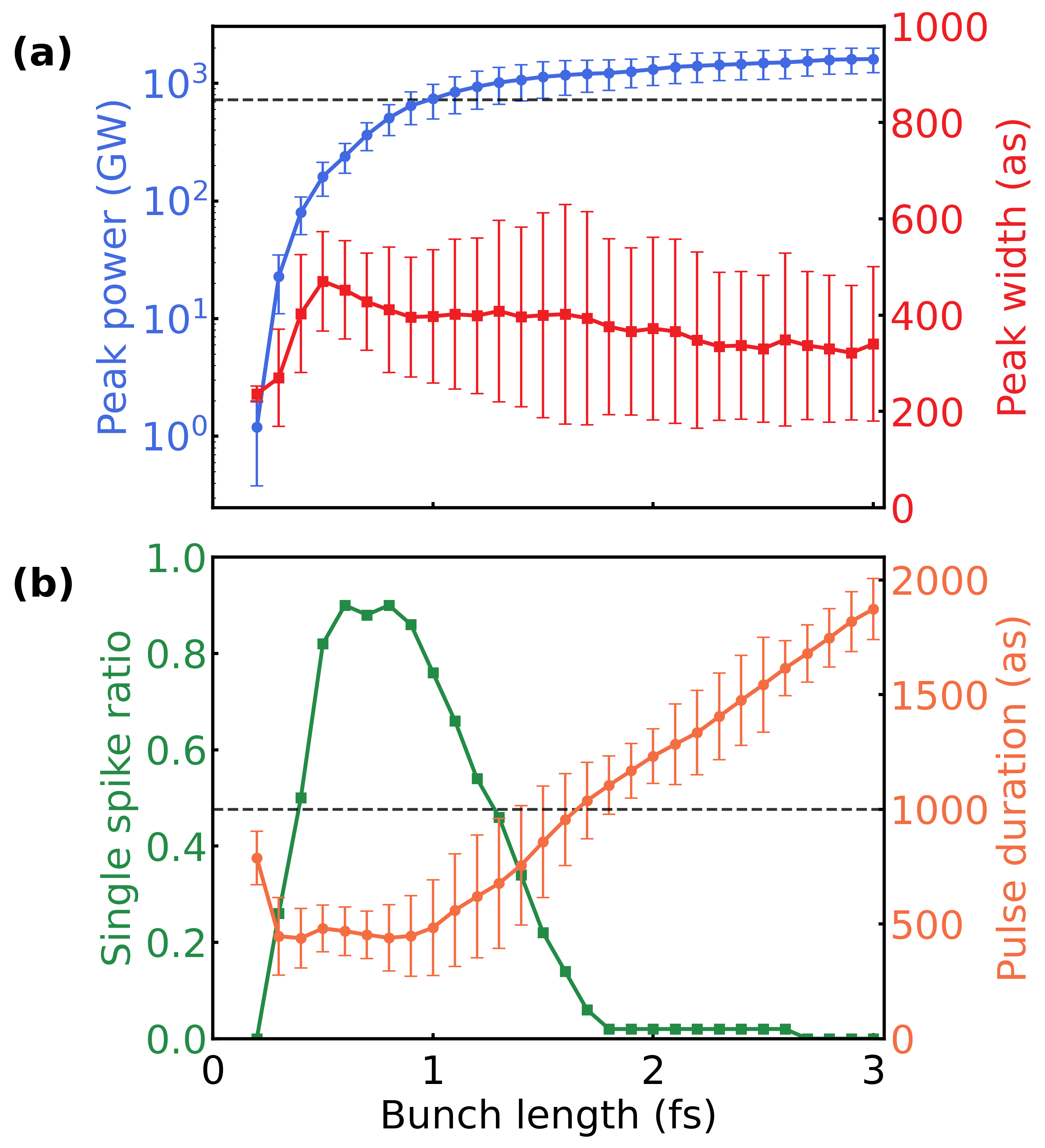}
\caption{Dependence of the FEL pulse properties on the electron-bunch length in the soft X-ray regime at $1000~\mathrm{eV}$. (a) Peak power and peak width. The horizontal black line indicates the saturation power calculated from the Ming Xie fitting formulas~\cite{xie1995design}. (b) Pulse duration and single-spike ratio. The horizontal black line marks a pulse duration of $1~\mathrm{fs}$. Error bars denote the shot-to-shot standard deviation obtained from 50 simulations at each bunch length.}
\label{Fig:part1_bunch_length_mean_peak_power_soft}
\end{figure}

A similar dependence on bunch length is observed in the soft X-ray case, as shown in Fig.~\ref{Fig:part1_bunch_length_mean_peak_power_soft}. The electron-beam parameters are kept the same as in the hard X-ray case, while the undulator period is increased to $70~\mathrm{mm}$ to cover a broader soft X-ray range, with the photon energy set to $1000~\mathrm{eV}$. Simulations are again carried out with GENESIS~1.3 (version~4) for eight undulator segments. The corresponding Pierce parameter is $\rho=2.69\times10^{-3}$, and the saturation power estimated from the Ming Xie fitting formulas is $723~\mathrm{GW}$. As in the hard X-ray case, the mean peak power increases with bunch length and begins to exceed the calculated saturation power at around $1~\mathrm{fs}$. It rises above $1~\mathrm{TW}$ for bunch lengths around $1.3~\mathrm{fs}$, while the 90th percentile of the peak-power distribution reaches $1.3~\mathrm{TW}$. At the same time, the pulse duration remains below $1~\mathrm{fs}$ for bunch lengths shorter than about $1.6~\mathrm{fs}$. A notable difference from the hard X-ray case is that the single-spike ratio remains substantially higher over a much wider bunch-length range. For example, even at a bunch length of $1~\mathrm{fs}$, the single-spike ratio is still about $76\%$. Physically, this behavior is consistent with the longer cooperation length in the soft X-ray regime, so that a given bunch length contains fewer cooperation lengths and is therefore less likely to support the development of multiple independent spikes. In addition to a sufficiently long electron bunch, a sufficiently long radiator is required for the radiation to enter the post-saturation superradiant regime. The influence of the available radiator length is further examined in Appendix~\ref{app:radiator_length}.

\section{Effects of other beam properties}

From the perspective of post-saturation superradiance, the discussion above establishes how high peak power attosecond pulse generation in a single-stage SASE XFEL depends on the electron bunch length. We next examine how other beam properties, including slice energy spread, slice emittance, energy chirp, and transverse beam tilt, modify this behavior.

\begin{figure}[tbp]
\centering
\includegraphics[width=0.9\linewidth]{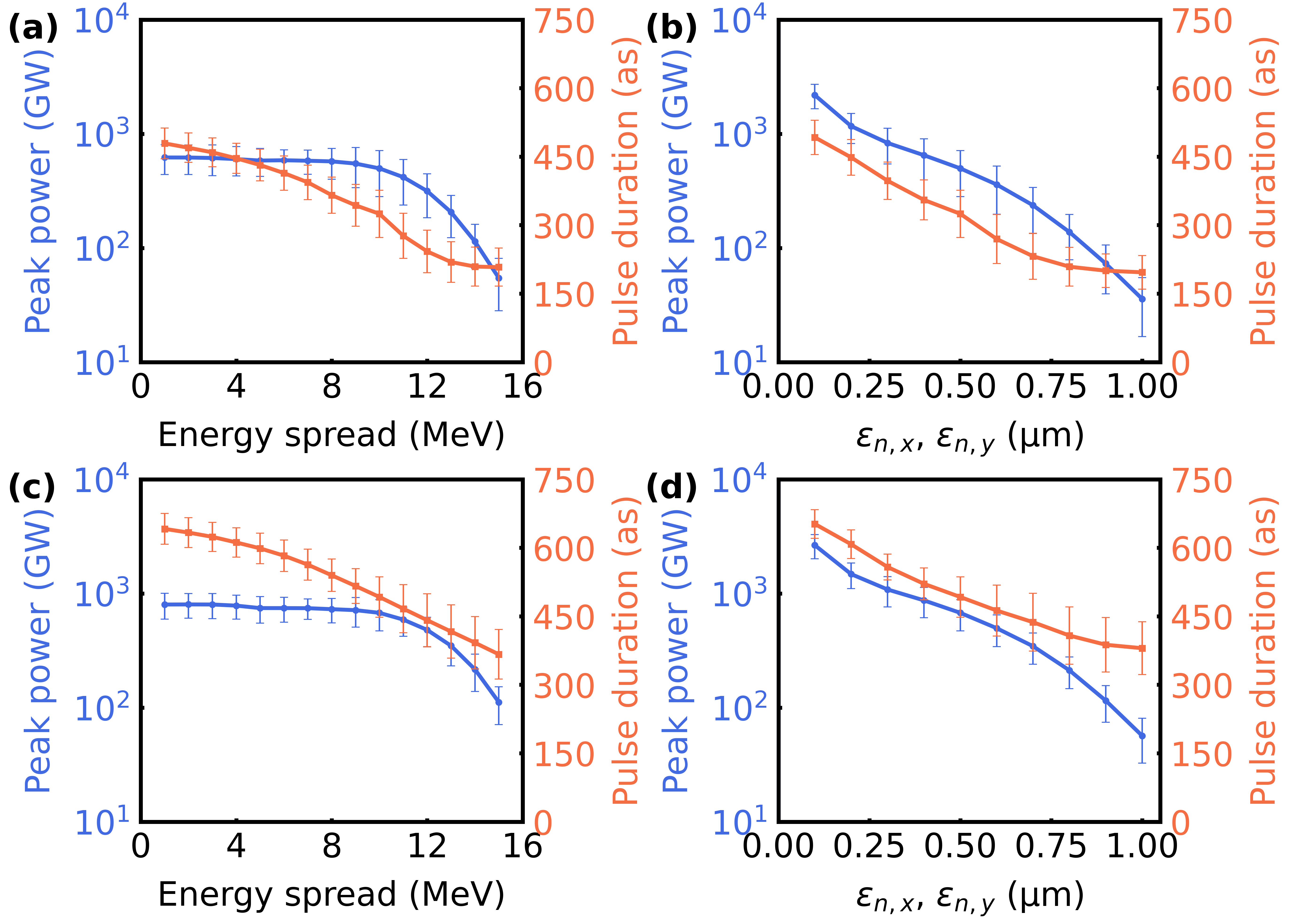}
\caption{Dependence of the mean peak power and pulse duration on the slice energy spread for bunch lengths of $650~\mathrm{as}$ (a) and $1~\mathrm{fs}$ (c), and on the normalized transverse emittance for bunch lengths of $650~\mathrm{as}$ (b) and $1~\mathrm{fs}$ (d). Error bars denote the shot-to-shot standard deviation obtained from 50 simulations at each parameter value.}
\label{Fig:part3_energy_spread_emit}
\end{figure}

\begin{figure}[tbp]
\centering
\includegraphics[width=0.9\linewidth]{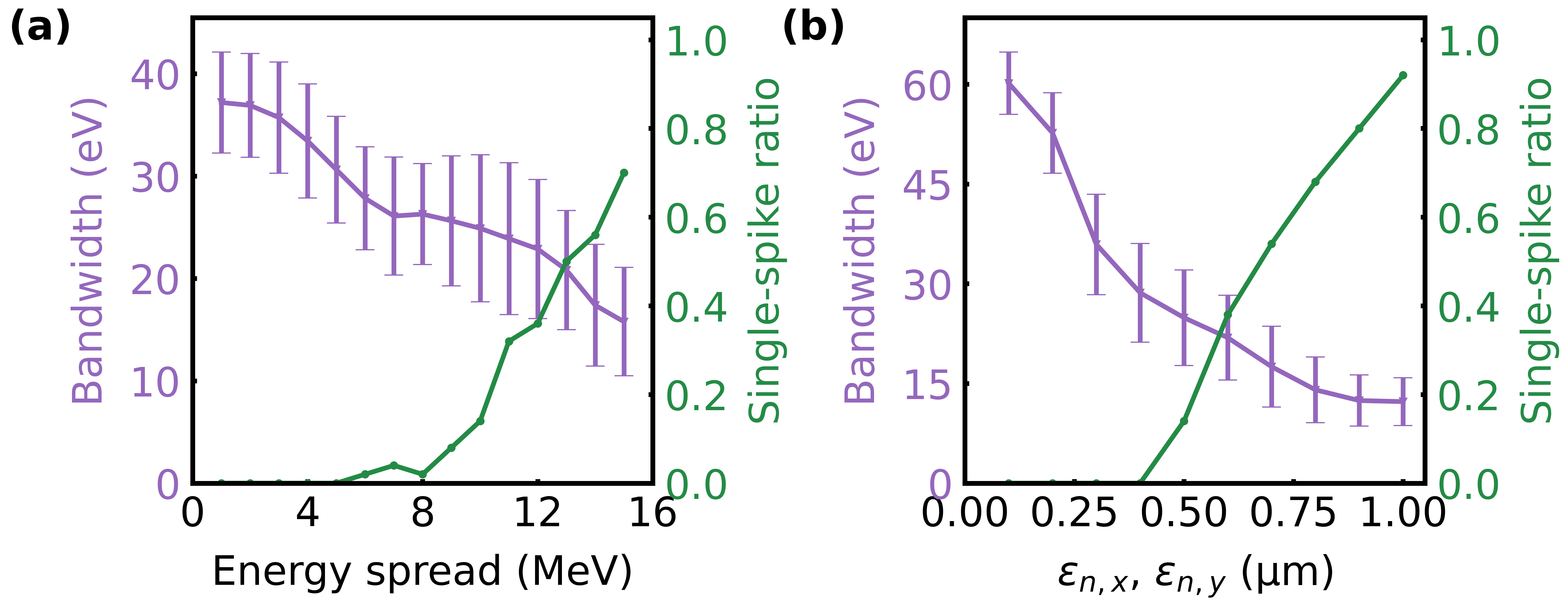}
\caption{Spectral bandwidth and single-spike ratio as functions of the slice energy spread (a) and normalized transverse emittance (b) for the $650~\mathrm{as}$ bunch. Error bars denote the shot-to-shot standard deviation obtained from 50 simulations at each parameter value.}
\label{Fig:part3_add_energy_spread_emit}
\end{figure}
We first analyze the effects of the slice energy spread and the normalized transverse emittance. Time-dependent simulations are performed with GENESIS~1.3 (version~4) for bunch lengths of $650~\mathrm{as}$ and $1~\mathrm{fs}$, with all other beam and undulator parameters kept fixed to generate pulses with photon energy $8700~\mathrm{eV}$. For the slice energy spread, the range is taken from $1$ to $15~\mathrm{MeV}$. For the emittance, we set $\varepsilon_{n,x}=\varepsilon_{n,y}$ and vary the normalized emittance from $0.1$ to $1.0~\mathrm{\mu m}$. For each parameter value, 50 shot-noise simulations are carried out.

\begin{figure*}[t]
\centering
\includegraphics[width=1\linewidth]{figure_9.png}
\caption{Temporal power profiles and longitudinal evolution for a $650~\mathrm{as}$ electron bunch with different slice energy spreads. The upper panel shows the normalized temporal power profiles of 50 shots at the undulator exit for slice energy spreads of $8$, $10$, $12$, and $14~\mathrm{MeV}$. The lower panels show the corresponding mean peak power (a), mean pulse duration (b), and temporal single-spike ratio (c) as functions of the undulator distance. The results in the lower panels are obtained from the same 50 shots for each slice energy spread. Error bars in panels (a) and (b) denote the shot-to-shot standard deviations.}
\label{Fig:part3_energy_spread_shot}
\end{figure*}

\begin{figure}[!htb]
\centering
\includegraphics[width=0.9\linewidth]{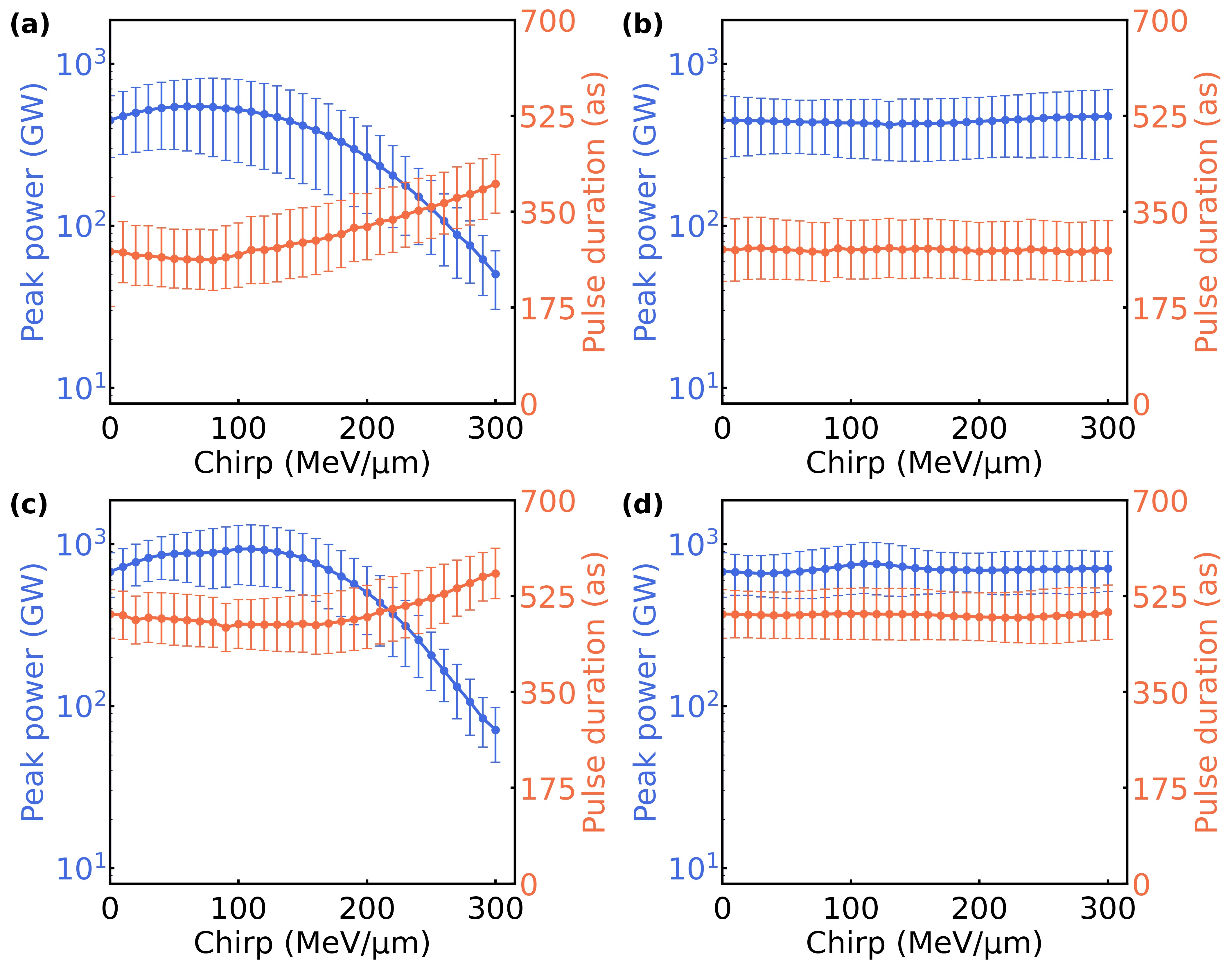}
\caption{Mean peak power and pulse duration as functions of the linear energy chirp for electron-bunch lengths of $650~\mathrm{as}$ and $1~\mathrm{fs}$. Panels (a) and (b) show the untapered and chirp-taper compensation cases for $650~\mathrm{as}$, respectively. Panels (c) and (d) show the corresponding untapered and chirp-taper compensation cases for $1~\mathrm{fs}$, respectively. Error bars denote the shot-to-shot standard deviation obtained from 200 simulations for the $650~\mathrm{as}$ case and 50 simulations for the $1~\mathrm{fs}$ case at each parameter value.}
\label{Fig:part2_length_single_chirp}
\end{figure}
The results are summarized in Fig.~\ref{Fig:part3_energy_spread_emit}. For both bunch lengths, the mean peak power decreases as either the slice energy spread or the normalized transverse emittance increases. For the case with a bunch length of $650~\mathrm{as}$, even at a slice energy spread of $14~\mathrm{MeV}$, the mean peak power remains about $80~\mathrm{GW}$. For the case with a bunch length of $1~\mathrm{fs}$, at the same slice energy spread, the mean peak power remains about $200~\mathrm{GW}$. The dependence on emittance is more sensitive. For both the $650~\mathrm{as}$ and $1~\mathrm{fs}$ cases, the mean peak power exceeds $2~\mathrm{TW}$ when $\varepsilon_n=0.1~\mathrm{\mu m}$, although such a small emittance is not realistic at present. For a transverse emittance as large as $0.8~\mathrm{\mu m}$, the mean peak power still remains $116~\mathrm{GW}$ for the $650~\mathrm{as}$ case, while for the $1~\mathrm{fs}$ case the mean peak power remains $212~\mathrm{GW}$. This behavior can be understood from the fact that smaller slice energy spread and emittance favor higher gain in the exponential regime~\cite{xie1995design}. They also shorten the gain length and, correspondingly, the cooperation length, allowing the radiation to reach the post-saturation regime within a shorter undulator distance and thus increasing the peak power. 

At the same time, the pulse duration also decreases as the slice energy spread or emittance increases. For the case with a bunch length of $650~\mathrm{as}$, the pulse duration approaches $200~\mathrm{as}$ when the slice energy spread is $14~\mathrm{MeV}$ or the normalized emittance is $0.8~\mathrm{\mu m}$. This trend can be understood within the same physical picture. As the slice energy spread or the normalized transverse emittance decreases, the cooperation length becomes shorter, allowing more spikes to grow to high power within the electron beam. Although this enhances the peak power, the development of additional spikes broadens the overall pulse envelope. The relationship between these trends and the cooperation length is discussed further in Appendix~\ref{app:energy_spread_emittance}.

It is also worth examining the impact of slice energy spread and emittance on the single-spike ratio. Here we focus on the case with a bunch length of $650~\mathrm{as}$, and the results are shown in Fig.~\ref{Fig:part3_add_energy_spread_emit}. The single-spike ratio increases substantially with increasing slice energy spread and normalized transverse emittance. For a normalized emittance of $0.8~\mathrm{\mu m}$, the single-spike ratio reaches 68\%, while for a slice energy spread of $14~\mathrm{MeV}$, it reaches 56\%. To illustrate this trend more clearly and examine its evolution along the undulator, we take the slice energy spread as an example. The upper panel of Fig.~\ref{Fig:part3_energy_spread_shot} shows the normalized temporal power profiles of 50 shots at the undulator exit for slice energy spreads of $8$, $10$, $12$, and $14~\mathrm{MeV}$, while the lower panels show the corresponding mean peak power, mean pulse duration, and temporal single-spike ratio along the undulator. The mean peak-power evolution shows that the cases with smaller slice energy spread enter the post-saturation regime within a shorter undulator distance, consistent with the discussion above. For all four slice energy spreads, the single-spike ratio first increases to a maximum and then decreases with further propagation, consistent with the behavior shown in Fig.~\ref{Fig:part1_650as_jitter_analysis}. As the slice energy spread increases, the maximum single-spike ratio becomes higher and is reached farther downstream, reflecting the increases in the cooperation length and gain length, respectively. The minimum pulse duration also occurs farther downstream. These results suggest useful tuning knobs for attosecond pulse generation through control of the slice energy spread and transverse emittance. If a higher single-spike ratio is prioritized, a somewhat larger slice energy spread or transverse emittance may be favorable, although this comes at the cost of reduced peak power. In practice, the slice energy spread may, for example, be controlled using a laser heater~\cite{PhysRevSTAB.7.074401}.

In attosecond pulse-generation schemes, whether laser-based or accelerator-based, an energy chirp is typically required to drive the compression needed to form a narrow high-current spike. In laser-based schemes, the chirp is usually introduced through energy modulation by an external laser. In accelerator-based schemes, it can be generated by off-crest acceleration or by collective effects such as longitudinal space charge~\cite{LSC} and coherent synchrotron radiation (CSR)~\cite{SALDIN1997373}. To analyze the impact of energy chirp on attosecond XFEL pulse generation, we consider electron beams with bunch lengths of $650~\mathrm{as}$ and $1~\mathrm{fs}$, while keeping the other parameters unchanged for a photon energy of $8700~\mathrm{eV}$. A linear positive chirp is imposed such that the beam head has higher energy, and the chirp strength is defined as the energy variation across the full bunch length per unit length. To make the dependence clearly visible, a wide chirp range from $0$ to $300~\mathrm{MeV}/\mu\mathrm{m}$ is considered.

Figures~\ref{Fig:part2_length_single_chirp}(a) and (c) show the untapered cases for $650~\mathrm{as}$ and $1~\mathrm{fs}$, respectively. For the $650~\mathrm{as}$ case, the mean peak power first increases with chirp and reaches about $546~\mathrm{GW}$ at around $60~\mathrm{MeV}/\mu\mathrm{m}$, compared with about $476~\mathrm{GW}$ at zero chirp. For the $1~\mathrm{fs}$ case, the mean peak power similarly increases from about $676~\mathrm{GW}$ at zero chirp to about $933~\mathrm{GW}$ at $110~\mathrm{MeV}/\mu\mathrm{m}$. This behavior originates from slippage. During amplification, the radiation pulse moves toward the beam head, where the electron energy is higher. It therefore partially compensates for the radiation-induced energy loss of the electron beam and enhances the peak power~\cite{saldin2006self}.

As the chirp is increased further, the peak power begins to decrease. Even so, the peak power decreases only gradually over a fairly wide chirp range. For the $650~\mathrm{as}$ case, the peak power does not fall below the zero-chirp value of about $476~\mathrm{GW}$ until the chirp reaches about $130~\mathrm{MeV}/\mu\mathrm{m}$. Even for a much larger chirp of $220~\mathrm{MeV}/\mu\mathrm{m}$, the peak power remains about $204~\mathrm{GW}$. In the $1~\mathrm{fs}$ case, the peak power still remains about $106~\mathrm{GW}$ at a chirp of $280~\mathrm{MeV}/\mu\mathrm{m}$. This reduction in peak power can be understood from the fact that, for large chirp, the chirp-induced detuning across the bunch becomes comparable to the FEL gain bandwidth and reduces the FEL gain, so that the pulse cannot be fully amplified. At the same time, the pulse duration is also observed to increase with chirp for both bunch lengths.

We further analyze the effect of undulator tapering for chirped electron beams. For a linear chirp, the resonance condition for the taper profile $K(z)$ along the undulator is~\cite{saldin2006self}
\begin{equation}
\frac{dK}{dz} = -\frac{\left(1 + K_{0}^{2}/2\right)^{2}}{K_{0}} \frac{1}{\gamma_{0}^{3}} \, \frac{d\gamma}{c dt},
\label{eq:chirp_taper}
\end{equation}
where $K_{0}$ is the reference undulator parameter used in evaluating the taper condition, and $\gamma_{0}$ is the relativistic factor of a reference electron. In the simulations, the undulator parameter is kept constant within each segment and adjusted only from one segment to the next. Figures~\ref{Fig:part2_length_single_chirp}(b) and (d) show that, with this compensation, the peak power can be maintained at a high level even for large chirp, while the pulse duration also remains relatively short over a much wider chirp range. Even at a chirp of $300~\mathrm{MeV}/\mu\mathrm{m}$, the peak power remains about $470~\mathrm{GW}$ for the $650~\mathrm{as}$ case and about $700~\mathrm{GW}$ for the $1~\mathrm{fs}$ case.

\begin{figure}[!htb]
\centering
\includegraphics[width=0.9\linewidth]{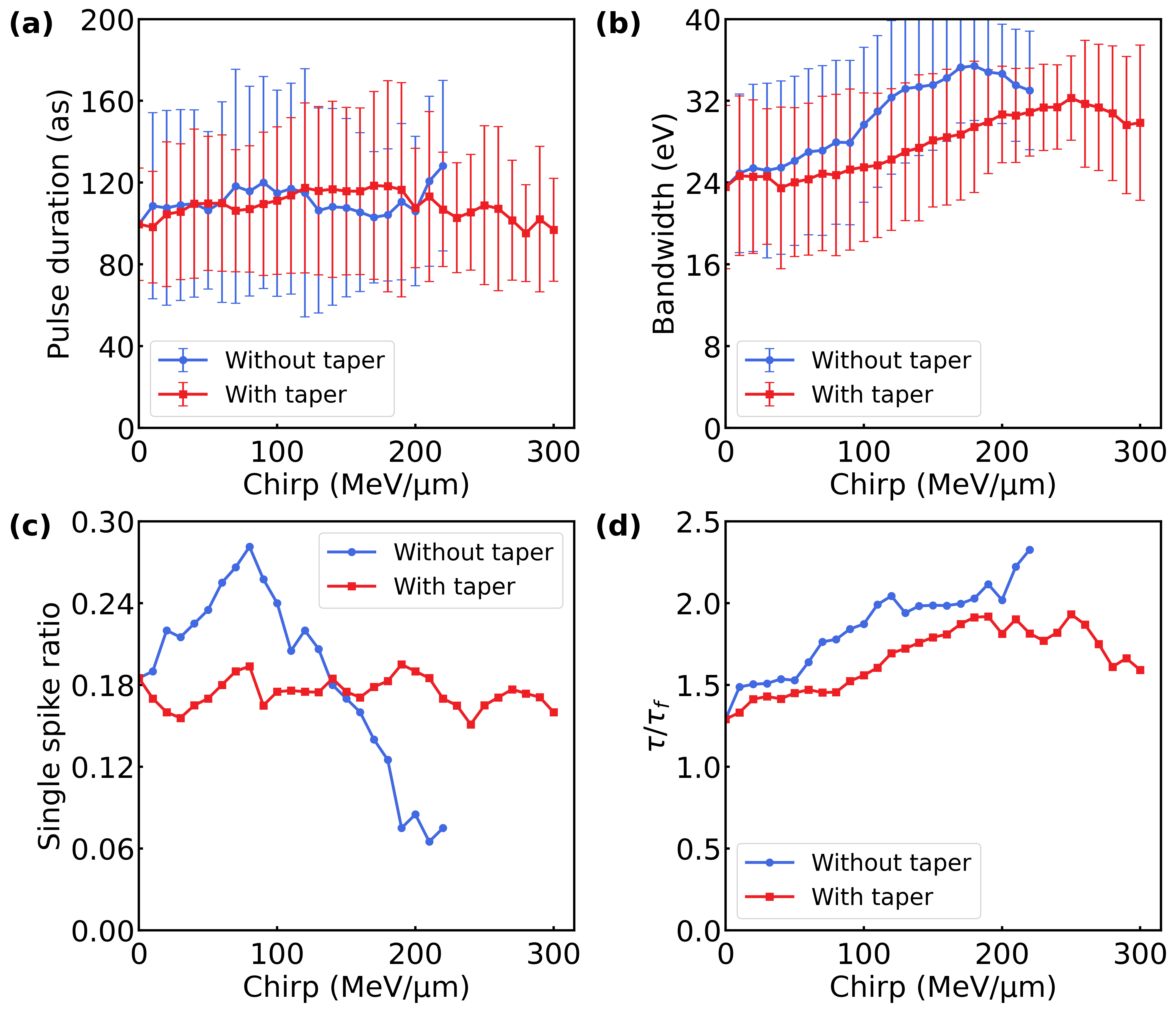}
\caption{Temporal and spectral properties of the attosecond XFEL pulse as functions of energy chirp for a bunch length of $650~\mathrm{as}$. Panels (a), (b), and (d) are evaluated from the temporal single-spike shots selected from 200 simulations at each chirp value. Panel (a) shows the pulse duration. Panel (b) shows the spectral bandwidth. Panel (c) shows the single-spike ratio obtained from all 200 simulations. Panel (d) shows the average value of $\tau/\tau_f$ over the selected temporal single-spike shots, where $\tau$ is the pulse duration and $\tau_f$ is the transform-limited reference duration.}
\label{Fig:part2_tbp_chirp}
\end{figure}

For the $650~\mathrm{as}$ bunch-length case, we perform 200 shot-noise simulations for each parameter set to examine the temporal and spectral properties in the presence of energy chirp and undulator tapering. As shown in Fig.~\ref{Fig:part2_tbp_chirp}(c), in the untapered case the single-spike ratio first increases slightly, from 18.5\% at zero chirp to 28\% at about $80~\mathrm{MeV}/\mu\mathrm{m}$, and then decreases as the chirp is increased further. Beyond about $220~\mathrm{MeV}/\mu\mathrm{m}$, the single-spike ratio falls below 5\%. In the chirp-taper case, the single-spike ratio remains above 18\%.

\begin{figure}[tbp]
\centering
\includegraphics[width=0.9\linewidth]{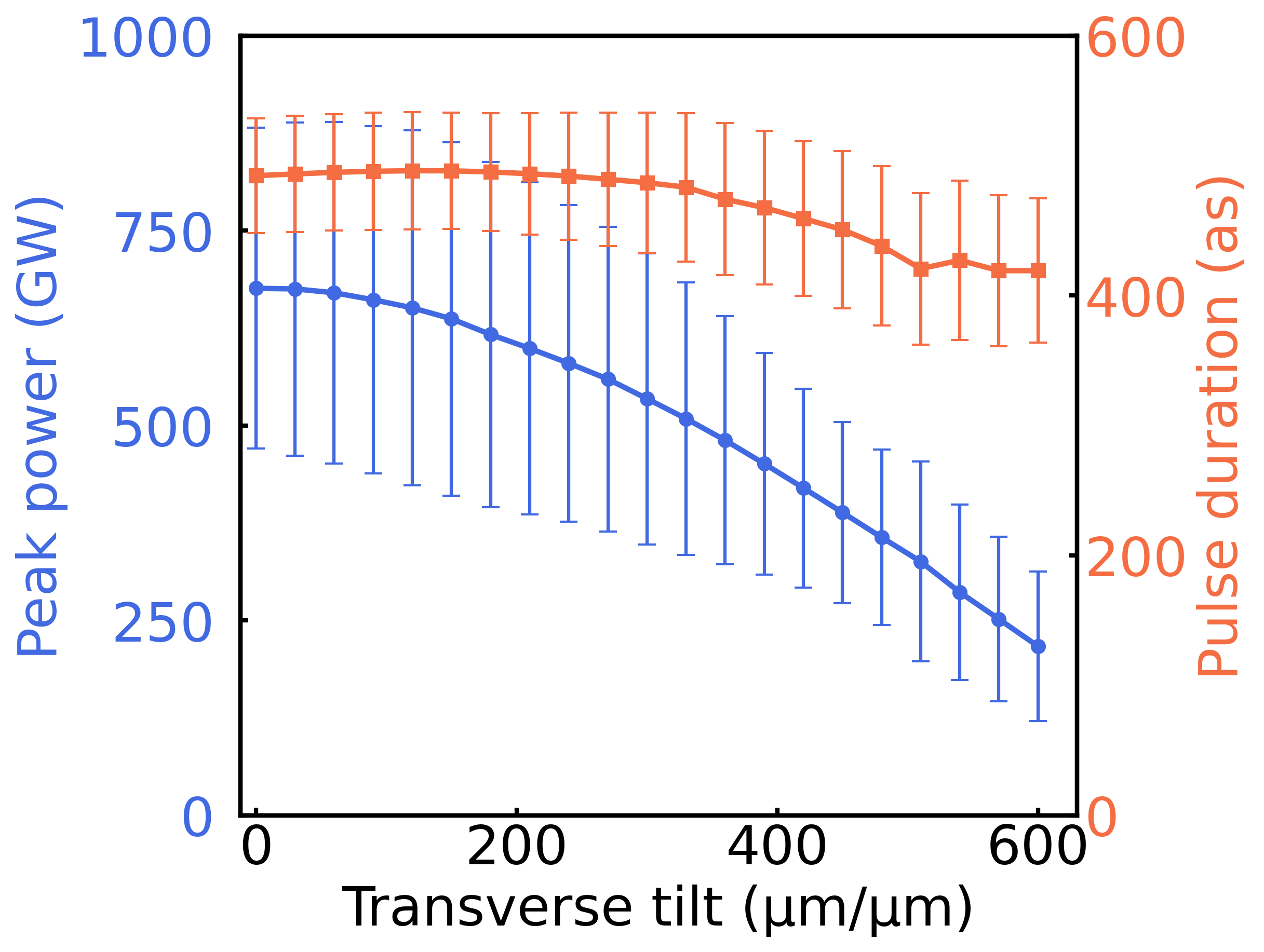}
\caption{Mean peak power and pulse duration as functions of the transverse tilt for an electron-bunch length of $1~\mathrm{fs}$. Error bars denote the shot-to-shot standard deviation obtained from 50 simulations at each tilt value.}
\label{Fig:part3_tilt}
\end{figure}

Because direct temporal characterization of hard X-ray attosecond XFEL pulses remains challenging, a common experimental practice is to use the spectral bandwidth to infer a reference pulse duration~\cite{Huang2017PRL}. Here, we consider the temporal single-spike cases selected from 200 shots for the $650~\mathrm{as}$ bunch-length case, for which a sufficient number of such shots is available. For a Gaussian pulse, the Fourier-limit reference duration is given by $\tau_f \approx 1.82/\Delta E~\mathrm{fs}$, where $\Delta E$ is the spectral bandwidth in eV. Figure~\ref{Fig:part2_tbp_chirp}(d) shows the average ratio $\tau/\tau_f$ over the selected temporal single-spike shots, where $\tau$ is the pulse duration. At zero chirp, this ratio is about 1.3 and increases gradually with chirp. For both the untapered and chirp-taper cases, it stays in the range of about 1.3 to 2 up to $180~\mathrm{MeV}/\mu\mathrm{m}$. For the chirp-taper case, it remains below 2 over the full chirp range up to $300~\mathrm{MeV}/\mu\mathrm{m}$.

Besides the electron beam properties discussed above, the transverse tilt of the electron beam can also affect FEL lasing~\cite{emma2004femtosecond,prat2015efficient}. Such a tilt may be introduced intentionally, for example by wakefields in a dechirper~\cite{Lutman2016NatPhoton}, by combining energy chirp with controlled dispersion in a dispersive section~\cite{Guetg2018PRL}, or by sextupole tuning in a bunch compressor to impose a quadratic tilt~\cite{PhysRevAccelBeams.23.030703}. It can also arise unintentionally from orbit distortions induced by CSR in bending systems~\cite{ob,dt,long:fel2022-tup44} or from residual transverse dispersion in an arc section.

To study the effect of transverse tilt, we apply a horizontal tilt at the undulator entrance for a $1~\mathrm{fs}$ bunch, with the tilt ranging from $0$ to $600~\mu\mathrm{m}/\mu\mathrm{m}$. Here, the tilt is defined as the horizontal offset per unit longitudinal position within the bunch, while all other parameters are kept unchanged for a photon energy of $8700~\mathrm{eV}$. Figure~\ref{Fig:part3_tilt} shows that the peak power decreases with increasing tilt. The pulse duration remains nearly unchanged up to about $200~\mu\mathrm{m}/\mu\mathrm{m}$, and then decreases as the tilt increases further. For a very large tilt of $600~\mu\mathrm{m}/\mu\mathrm{m}$, the peak power is about $200~\mathrm{GW}$ and the pulse duration is about $420~\mathrm{as}$. This trend indicates that a transverse tilt reduces the effective lasing window and thereby affects attosecond pulse generation in a manner similar to shortening the electron bunch.

\section{Design strategy for attosecond XFEL}

The analysis in the preceding sections establishes a physical framework for attosecond pulse generation from a short current spike, in which the electron-bunch length plays a central role. In practice, the bunch length can be tailored through charge control and advanced bunch compression, including manipulation of the linear momentum compaction $R_{56}$ and the second-order term $T_{566}$. The design considerations discussed below become particularly demanding in the hard X-ray regime, where the cooperation length is short relative to the current-spike length and single-spike operation competes with access to the post-saturation regime. In the soft X-ray regime, the longer cooperation length relaxes this tension, so that high peak power and single-spike behavior are more readily achieved over a wider parameter range, as illustrated in Fig.~\ref{Fig:part1_bunch_length_mean_peak_power_soft} and Fig.~\ref{Fig:radiator_length}. Different classes of attosecond X-ray experiments place different weights on three principal pulse properties, namely peak power, pulse duration, and single-spike probability. A given application typically benefits from improvement in all three, but the achievable trade-off depends on how the electron beam is shaped.

Experiments aimed at resolving coherent electronic wavepacket evolution, such as charge migration and attosecond photoionization, rely on a well-defined temporal envelope of the pump and probe X-ray pulses, since secondary structures can introduce ambiguity in the reconstruction of electronic dynamics. For such studies, operating with a relatively short electron bunch yields a substantially higher single-spike probability, which is important for clean pump-probe measurements. However, the short lasing window limits the attainable peak power.

A distinct operating mode is accessed by using a somewhat longer electron bunch, exemplified by the 650~as case studied above. Here the bunch is long enough for the dominant spike to enter the post-saturation regime, so that single-spike shots reach substantially higher peak power than in the short-bunch case. The trade-off is that such single-spike events occur with lower probability, around 20\% in our simulations, with the remaining shots developing secondary structures. Since attosecond XFEL experiments routinely rely on shot-by-shot spectral or temporal diagnostics, post-selection of single-spike events is possible. In this regime, peak powers approaching 1~TW and pulse durations below 70~as can be obtained on selected shots, at the cost of a reduced single-spike probability.

Nonlinear and multi-photon X-ray experiments, along with attosecond coherent diffractive imaging aimed at electronic damage-free structure determination, benefit from higher peak intensity on the sample while requiring sub-femtosecond pulse durations to outrun Auger decay and subsequent ionization cascades. Within these constraints, a longer electron bunch allows the dominant spike to evolve further into the post-saturation regime, reaching higher peak power, provided the pulse envelope remains sufficiently short. In this mode, single-spike events become rare, but the mean peak power averaged over all shots is maximized.

Applications that demand simultaneously high peak power, short pulse duration, and high single-spike probability motivate more complex schemes such as multi-stage amplification~\cite{tanaka2016using}. In such schemes, an optimized electron bunch in the first stage establishes a dominant spike with a high probability by limiting the effective lasing window, while one or more downstream stages further amplify this spike and can drive it into the post-saturation superradiant regime. Within the framework developed here, multi-stage amplification effectively extends the lasing window seen by the dominant spike while suppressing the growth of competing structures.

Across all of these operating modes, the electron-bunch length emerges as the primary control parameter linking beam physics to the target pulse format. Even in multi-stage configurations, the bunch length in the initial stage must be carefully chosen to balance single-spike probability and the peak power passed to the next stage. This in turn implies that high-precision measurement of the bunch length is an essential capability for attosecond XFEL operation~\cite{10.1117/1.AP.7.2.026002}.

Beyond bunch length, several additional beam parameters provide tuning knobs for tailoring the output pulse properties. Slice energy spread and transverse emittance can be deliberately manipulated to improve single-spike probability, which is useful when further reduction of the bunch length is constrained by beam dynamics or machine limitations. A moderate increase in slice energy spread via a laser heater promotes the formation of single-spike shots at the cost of reduced peak power. An appropriate positive energy chirp can further enhance the peak power. Looking further ahead, pushing peak currents beyond the tens-of-kA regime typical of current attosecond XFEL operation, potentially to $\sim$100~kA, would provide an additional route to higher peak power.

Realizing the precise beam manipulation strategies outlined above at their full potential is challenging at existing user facilities, which are typically designed for broad-wavelength, multi-mode operation. As attosecond science continues to develop, the demand for attosecond XFELs is expected to grow substantially, motivated by their broad wavelength coverage up to the hard X-ray regime and their high peak power. This motivates the development of dedicated attosecond XFEL facilities, for which two main routes can be envisioned. The first is a compact FEL design, similar in spirit to SACLA~\cite{sacla} and SwissFEL~\cite{swissFEL} but specifically optimized for attosecond operation. The second is based on a continuous-wave (CW) superconducting XFEL facility, which can support multiple undulator lines simultaneously. Recent progress in CW-XFELs~\cite{cwlcls2015,cwlcls2021,zhao2018shine,cwshine2023,
cws3fel,zhu2022inhibition} has opened the possibility of combining attosecond pulse generation with very high repetition rates, which is particularly attractive for applications such as imaging 
that outruns electronic damage and for high-repetition-rate nonlinear X-ray experiments. A possible dedicated attosecond facility design based on a CW superconducting accelerator is outlined in Appendix~\ref{sec:appendix_facility}.

\section{Conclusions}

In this work, we have developed a unified framework for high-power attosecond XFEL generation from a short current spike, extending and quantifying the role of post-saturation superradiant evolution previously highlighted in Ref.~\cite{yan2025attoshine}. Once this regime is reached, the dominant radiation spike continues to grow in peak power while becoming shorter in time. From this perspective, the electron-bunch length is the primary control parameter, since it determines how long the radiation remains overlapped with the electrons under slippage and therefore how far the pulse can evolve toward the post-saturation regime. This leads to a clear trade-off among peak power, pulse shortening, and single-spike probability. We further examined how slice energy spread, transverse emittance, energy chirp, and transverse beam tilt modify this picture. 

While post-saturation superradiance has previously been exploited through dedicated schemes such as seeded~\cite{Mirian2021}, chirp-taper~\cite{PhysRevAccelBeams.23.020702}, and two-stage amplification~\cite{franz2024terawatt}, we show that it already plays a central role in attosecond pulse generation from a standard single-stage SASE. The design principles established here therefore apply to any attosecond XFEL scheme relying on a short current spike, regardless of how the spike is generated or in which specific amplification configuration it reaches the post-saturation regime. In addition, all results reported here are obtained without a post-saturation taper, so that dedicated taper optimization is expected to provide an additional route toward even higher peak power.

The physical picture presented here provides general guidance for the design and optimization of electron beams in attosecond XFEL schemes. As new methods continue to emerge for producing ever shorter and more intense current spikes, the physics developed here will remain the common framework for understanding their performance. These results establish a physical basis for linking beam physics to the specific requirements of different classes of attosecond X-ray science, and provide a foundation for the design of next-generation dedicated attosecond XFEL facilities.

\section{Acknowledgments}

The authors thank Nanshun Huang from SARI, Toru Hara and Hitoshi Tanaka from the RIKEN SPring-8 Center, Ichiro Inoue from the University of Tokyo, and Gianluca Geloni from European XFEL for helpful discussions. This work was supported by the National Natural Science Foundation of China (12125508, 12541503), the National Key Research and Development Program of China (2024YFA1612104), the CAS Project for Young Scientists in Basic Research (YSBR-042), and the Shanghai Pilot Program for Basic Research--Chinese Academy of Sciences, Shanghai Branch (JCYJ-SHFY-2021-010). Jiawei Yan, Ye Chen, Winfried Decking, Marc Guetg, and Tianyun Long acknowledge support from DESY (Hamburg, Germany), a member of the Helmholtz Association (HGF), and from the European XFEL (Schenefeld, Germany). The numerical simulations were carried out using the ``MAXWELL'' computational resources operated at Deutsches Elektronen-Synchrotron DESY, Hamburg, Germany.

\appendix

\section{Influence of the Longitudinal Current Profile}
\label{sec:gaussian_profile}

The simulations presented in the main text use an idealized flat-top current distribution to isolate the effect of the electron-bunch length while keeping the other beam parameters fixed. To examine the influence of the longitudinal current profile, we performed additional simulations using a Gaussian current distribution with a FWHM duration of $650~\mathrm{as}$. The peak current was kept at $20~\mathrm{kA}$, and all other electron-beam and undulator parameters were the same as in the flat-top case.

\begin{figure}[tbp]
    \centering
    \includegraphics[width=1\linewidth]{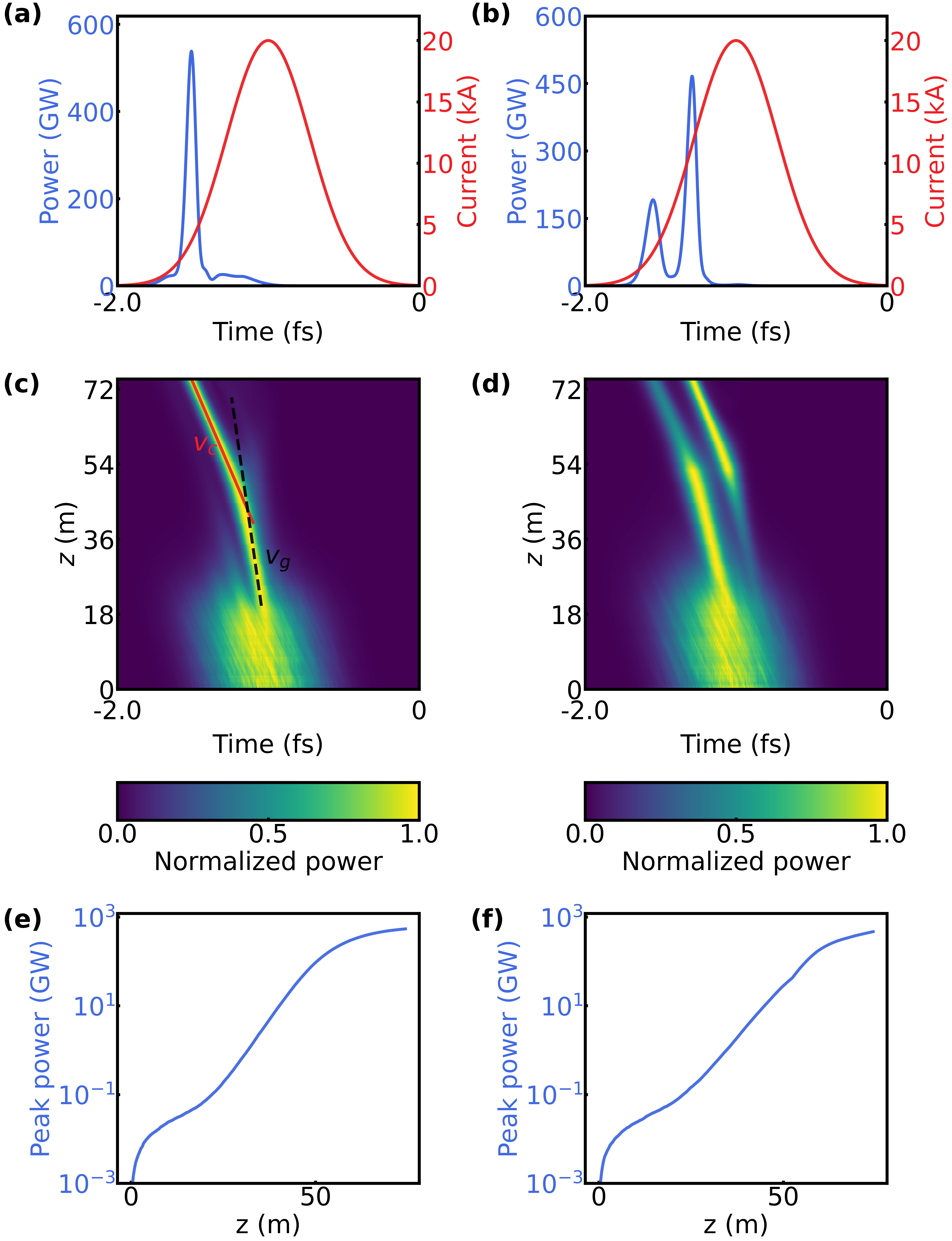}
    \caption{Results obtained using a Gaussian current distribution with a FWHM duration of $650~\mathrm{as}$. The left column shows a single-spike shot, and the right column shows a multi-spike shot. Panels (a) and (b) show the electron-current and radiation-power profiles at the undulator exit. Panels (c) and (d) show the two-dimensional evolution of the radiation power along the undulator. At each undulator position $z$, the power profile is normalized to its local maximum. Panels (e) and (f) show the evolution of the peak power along the undulator.
    }
    \label{Fig:gaussian_profile}
\end{figure}

A total of 50 shots were simulated. Eight shots were classified as single-spike, corresponding to a single-spike ratio of $16\%$. At the undulator exit, the mean peak power and mean pulse duration over all shots are $326~\mathrm{GW}$ and $330~\mathrm{as}$, respectively. Figure~\ref{Fig:gaussian_profile} shows a single-spike shot and a multi-spike shot. For the single-spike shot, the peak power exceeds $500~\mathrm{GW}$ and the pulse duration is approximately $100~\mathrm{as}$ at the undulator exit. The corresponding evolution along the undulator shows post-saturation behavior consistent with that observed in the flat-top case. Beyond the idealized electron-beam profiles considered here, Ref.~\cite{yan2025attoshine} presents detailed start-to-end simulations of attosecond XFEL pulse generation with an ultrashort electron bunch using the self-chirping scheme.

\section{Dependence on the Radiator Length}
\label{app:radiator_length}

\begin{figure*}[t]
\centering
\includegraphics[width=0.9\linewidth]{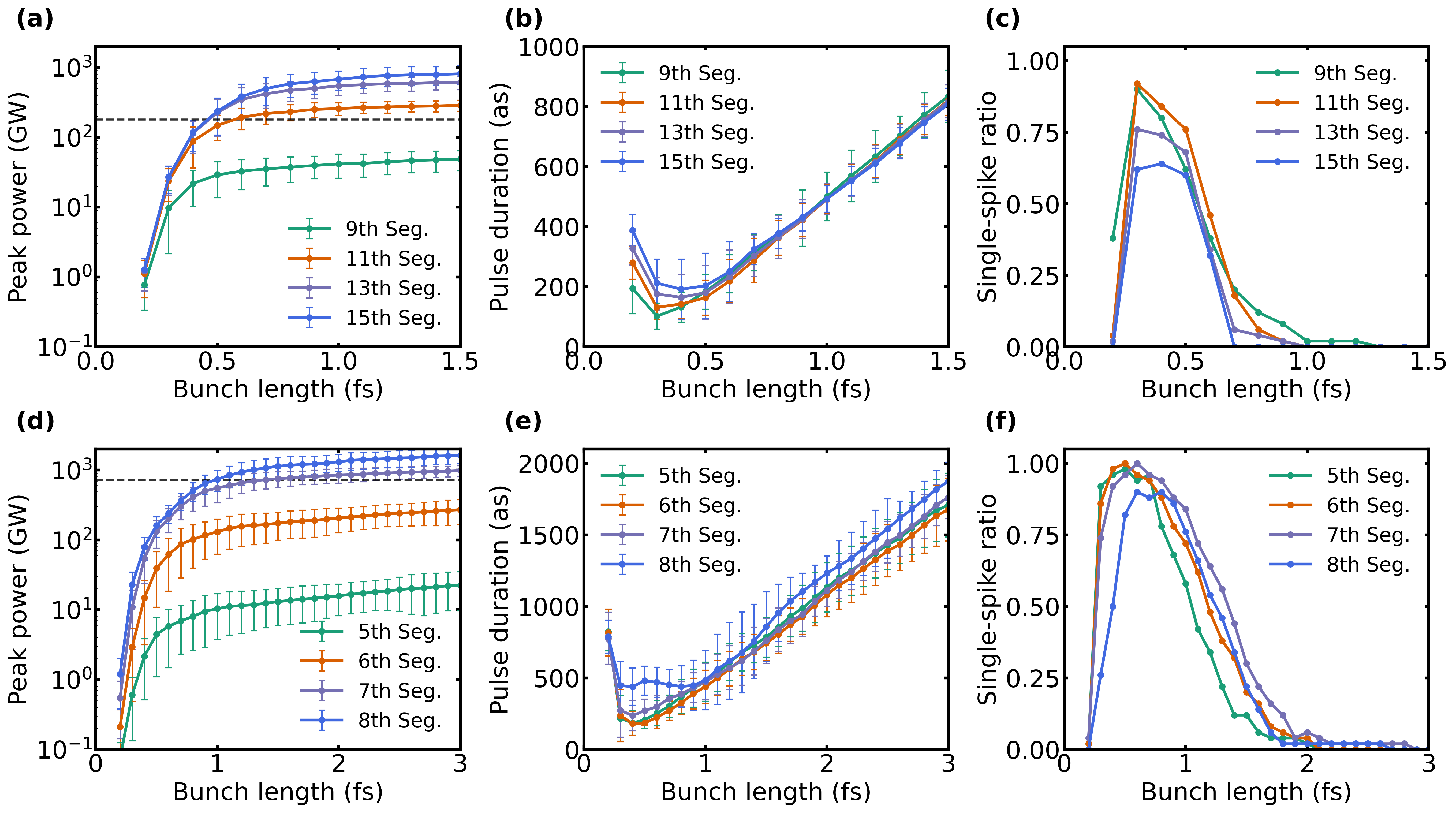}
\caption{Mean peak power, mean pulse duration, and temporal single-spike ratio as functions of the electron-bunch duration at the exits of selected undulator segments. Panels (a)--(c) show the hard X-ray results at the exits of the 9th, 11th, 13th, and 15th undulator segments, while panels (d)--(f) show the soft X-ray results at the exits of the 5th, 6th, 7th, and 8th undulator segments. The dashed lines in panels (a) and (d) indicate the saturation powers estimated from the Ming Xie fitting formulas. Each point is obtained from 50 shots, and the error bars in panels (a), (b), (d), and (e) denote the shot-to-shot standard deviations.}
\label{Fig:radiator_length}
\end{figure*}

The results presented in the main text are obtained at the exits of the 15th undulator segment for the hard X-ray case and the 8th undulator segment for the soft X-ray case, corresponding to active undulator lengths of $75$ and $40~\mathrm{m}$, respectively. Using the definition of $l_g$ introduced in the main text, the three-dimensional gain length is calculated as $l_{g,\mathrm{3D}}=(1+\Lambda)l_g$, where $\Lambda$ is obtained from the Ming Xie fitting formulas~\cite{xie1995design}. For the hard X-ray parameters used in the main text, the three-dimensional gain length is approximately $4.85~\mathrm{m}$, so the $75~\mathrm{m}$ radiator corresponds to about $15.5$ gain lengths. For the soft X-ray parameters, the three-dimensional gain length is approximately $2.67~\mathrm{m}$, so the $40~\mathrm{m}$ radiator corresponds to about $15.0$ gain lengths. Since a sufficiently long radiator is required for the radiation to enter and evolve in the post-saturation superradiant regime, we further examine the dependence of the results on the available radiator length.

Figure~\ref{Fig:radiator_length} shows the mean peak power, mean pulse duration, and temporal single-spike ratio as functions of the electron-bunch duration at the exits of selected undulator segments. For the hard X-ray case, the results are evaluated at the exits of the 9th, 11th, 13th, and 15th undulator segments, while the soft X-ray results are evaluated at the exits of the 5th, 6th, 7th, and 8th undulator segments. The overall dependence on the electron-bunch duration remains similar at the different undulator-segment exits. The mean peak power increases with the bunch duration, whereas the temporal single-spike ratio first increases and then decreases. The mean pulse duration correspondingly first decreases and then increases. To illustrate the influence of the radiator length at specific bunch durations, we consider a $500~\mathrm{as}$ electron bunch in the hard X-ray case and a $1~\mathrm{fs}$ electron bunch in the soft X-ray case. For the $500~\mathrm{as}$ hard X-ray case, the mean peak powers at the exits of the 9th, 11th, 13th, and 15th undulator segments are approximately $29$, $148$, $227$, and $236~\mathrm{GW}$, respectively, while the corresponding single-spike ratios are $62\%$, $76\%$, $68\%$, and $60\%$. The highest single-spike ratio is obtained at the exit of the 11th undulator segment, whereas additional segments increase the peak power but reduce the single-spike ratio. For the $1~\mathrm{fs}$ soft X-ray case, the mean peak powers at the exits of the 5th, 6th, 7th, and 8th undulator segments are approximately $10$, $131$, $552$, and $739~\mathrm{GW}$, while the corresponding single-spike ratios are $58\%$, $72\%$, $84\%$, and $76\%$. The highest single-spike ratio is obtained at the exit of the 7th undulator segment, whereas the 8th segment further increases the peak power at the cost of a lower single-spike ratio. These results show that the radiator length should be optimized according to the desired output. For a given electron-bunch length, an appropriate reduction in the radiator length can increase the single-spike ratio, but this comes at the cost of reduced peak power.

\section{Cooperation Length for Different Slice Energy Spreads and Transverse Emittances}
\label{app:energy_spread_emittance}

To further examine the effects of slice energy spread and transverse emittance, we express the results presented in Fig.~\ref{Fig:part3_energy_spread_emit} in terms of the cooperation time. Following the three-dimensional gain-length correction defined above, the corrected cooperation length and cooperation time are defined as $l_{c,\mathrm{3d}}=(1+\Lambda)l_c$ and $\tau_{c,\mathrm{3d}}=l_{c,\mathrm{3d}}/c$, respectively.

\begin{figure}[tbp]
\centering
\includegraphics[width=1\linewidth]{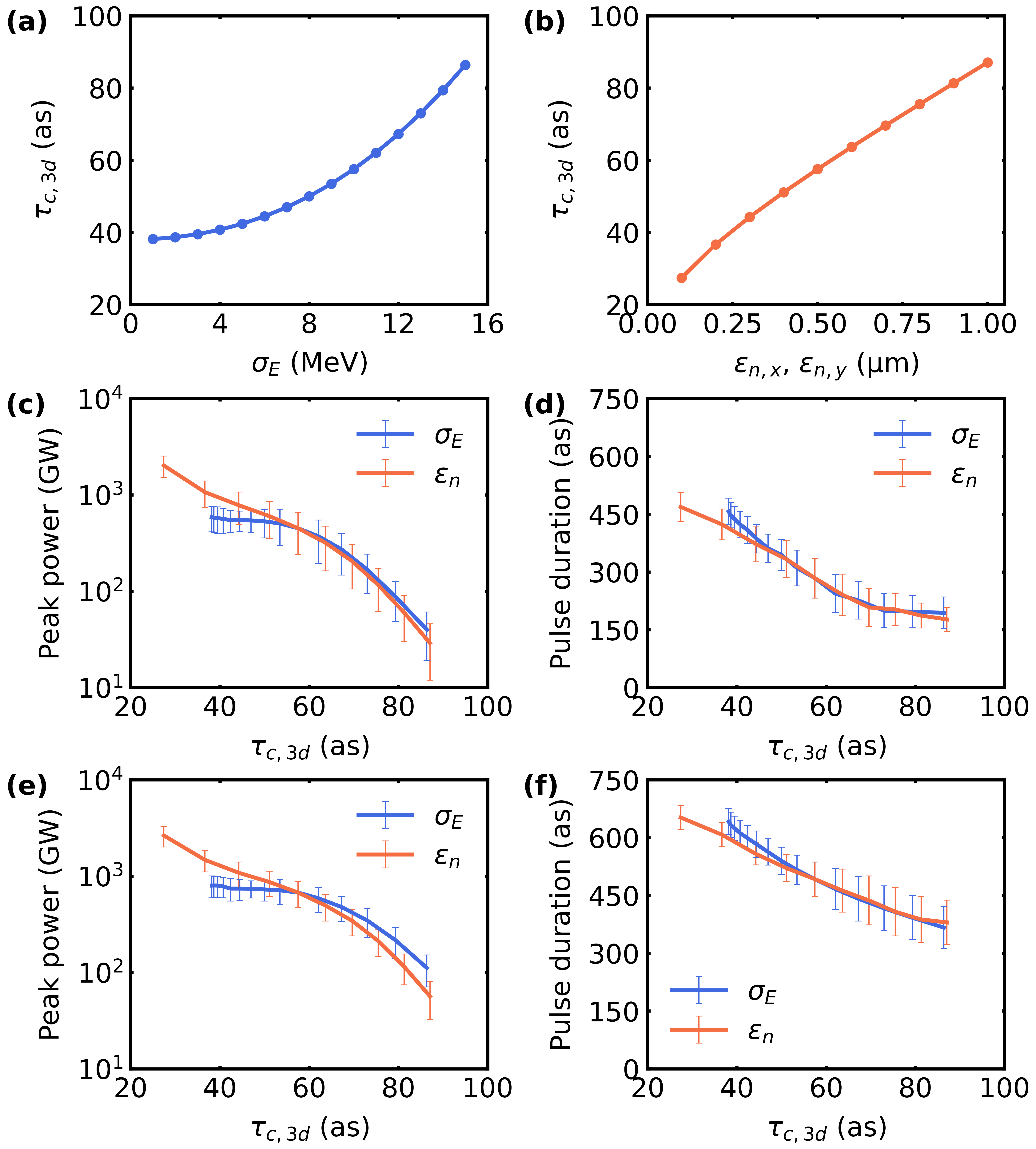}
\caption{Cooperation time and FEL output for different slice energy spreads and normalized transverse emittances. Panels (a) and (b) show $\tau_{c,\mathrm{3d}}$ as a function of the slice energy spread and normalized transverse emittance, respectively. Panels (c) and (d) show the mean peak power and mean pulse duration as functions of $\tau_{c,\mathrm{3d}}$ for a bunch duration of $650~\mathrm{as}$, while panels (e) and (f) show the corresponding results for a bunch duration of $1~\mathrm{fs}$. Blue and orange curves represent results obtained by varying the slice energy spread and normalized transverse emittance, respectively. Error bars indicate the shot-to-shot standard deviations from 50 simulations.}
\label{Fig:corrected_cooperation_time}
\end{figure}

Figure~\ref{Fig:corrected_cooperation_time} shows the calculated dependence of $\tau_{c,\mathrm{3d}}$ on the slice energy spread and normalized transverse emittance in panels (a) and (b), respectively. As either the slice energy spread or transverse emittance increases, the three-dimensional gain length and the corresponding cooperation time increase. For the standard hard-X-ray parameters $\sigma_E=10~\mathrm{MeV}$ and $\varepsilon_n=0.5~\mathrm{\mu m}$, the calculation gives $\tau_{c,\mathrm{3d}}\approx57.5~\mathrm{as}$. Panels (c) and (d) present the mean peak power and mean pulse duration as functions of $\tau_{c,\mathrm{3d}}$ for an electron-bunch duration of $650~\mathrm{as}$, while panels (e) and (f) show the corresponding results for an electron-bunch duration of $1~\mathrm{fs}$. For both bunch lengths, the mean peak power and mean pulse duration generally decrease as $\tau_{c,\mathrm{3d}}$ increases for variations in either the slice energy spread or the transverse emittance. These consistent overall trends show that the cooperation time provides a useful physical scale for analyzing the influence of electron-beam quality on the FEL output.

\begin{figure*}[t]
\centering
\includegraphics[width=1\linewidth]{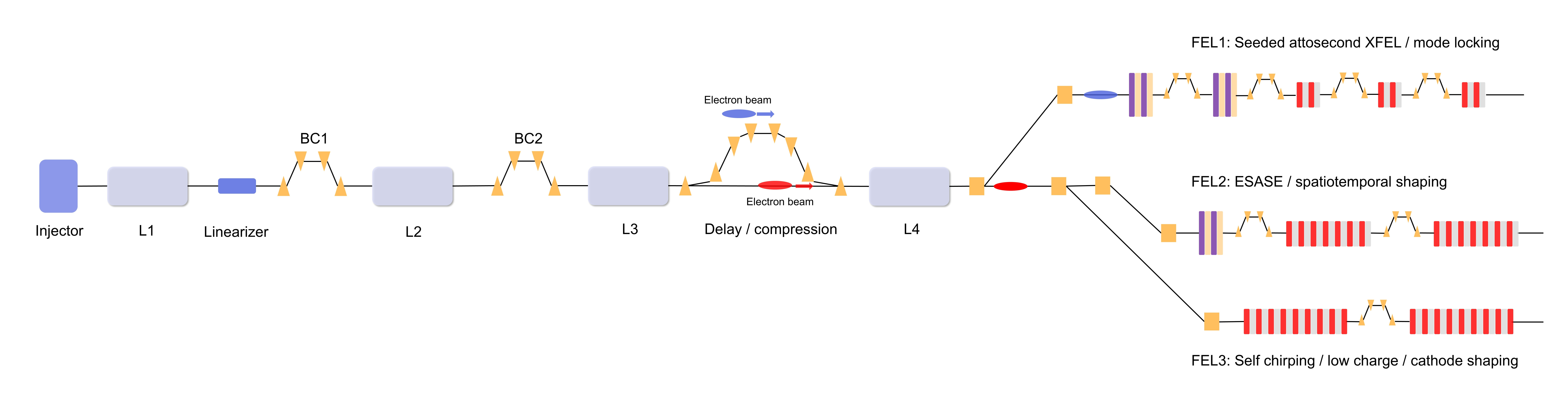}
\caption{Schematic layout of a dedicated attosecond XFEL facility. The injector, linearizer, linac sections (L1--L4), and bunch compressors provide the main beam acceleration and compression. After L3, a delay and compression system enables multienergy operation and additional bunch compression before L4~\cite{PhysRevAccelBeams.22.090701}. After L4, the beam is distributed to different FEL lines through switchyard sections, which can serve both as transport lines and as active compression sections. FEL1 is intended for external seeding and mode-locking operation, FEL2 for ESASE-based and spatiotemporal-shaping modes, and FEL3 for self-chirping, low-charge/high-charge operation, and high-photon-energy or high-pulse-energy attosecond generation.}
\label{Fig:part4_dedicated_attosecond}
\end{figure*}

\begin{figure}[tbp]
\centering
\includegraphics[width=1\linewidth]{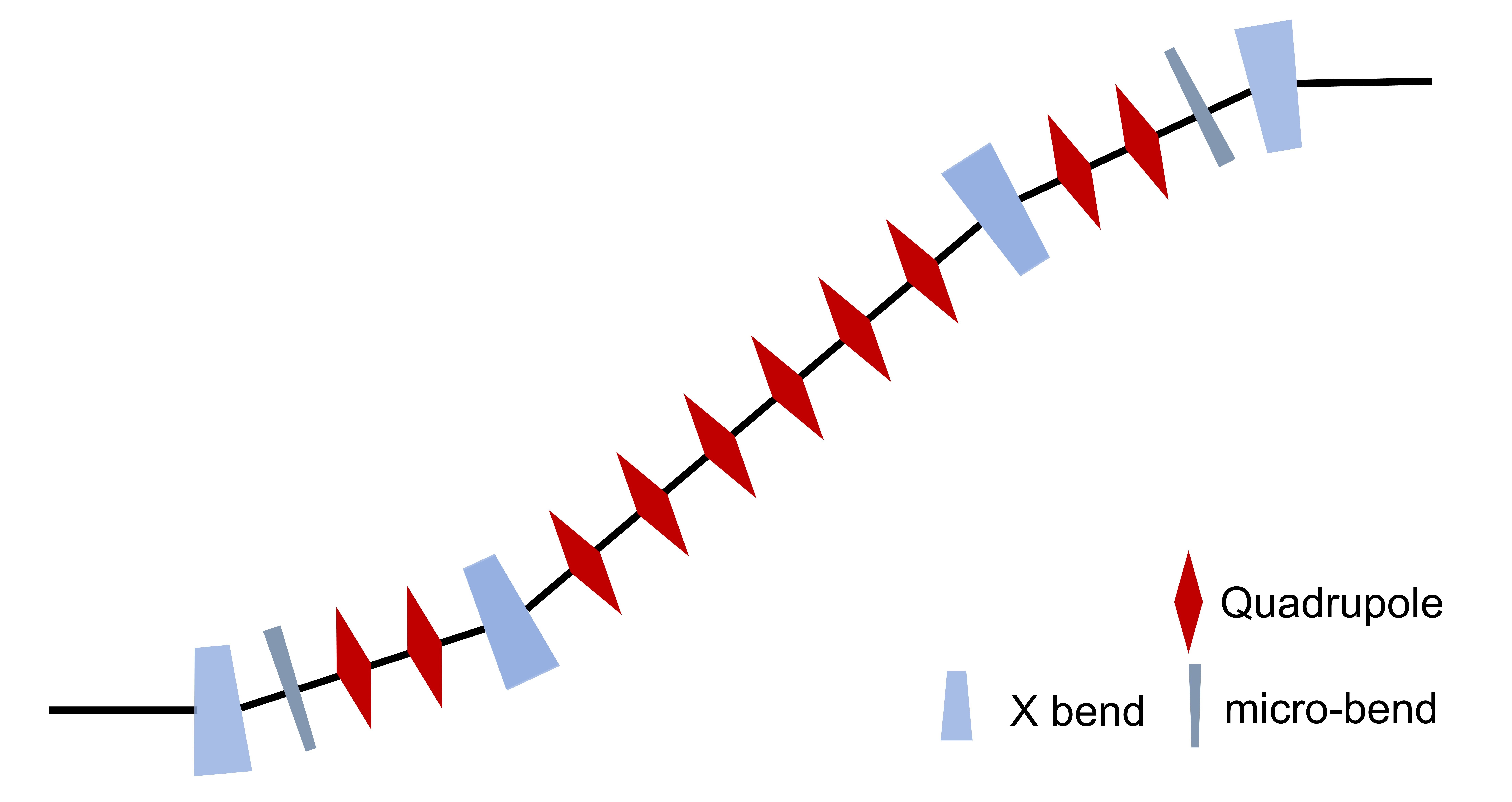}
\includegraphics[width=1\linewidth]{appendix_figure_5b.png}
\caption{Conceptual design of the arc section. The section consists of two DBAs and additional micro-bends, allowing flexible control of the longitudinal dispersion. Panels (a), (b), and (c) show representative optics and the corresponding $R_{56}$ distributions for the cases of zero, positive, and negative $R_{56}$, respectively, while preserving the achromatic condition.}
\label{Fig:part4_arc_design}
\end{figure}

\section{Design of a Dedicated CW Attosecond XFEL Facility}
\label{sec:appendix_facility}

The results presented in the main text show that precise control of the electron-beam parameters, including the bunch length, peak current, slice energy spread, and transverse emittance, is essential for optimizing attosecond FEL performance. A dedicated attosecond XFEL facility should therefore provide flexible control of these electron-beam parameters, together with the bunch charge and electron-beam energy, as well as multiple FEL lines designed to meet different user requirements and support future attosecond operating modes.  In this Appendix, we outline one possible implementation based on a CW superconducting accelerator, as illustrated in Fig.~\ref{Fig:part4_dedicated_attosecond}. High-repetition-rate electron bunches are generated in the photoinjector section, which can support both low-charge and high-charge operation. A laser heater can be incorporated downstream of the injector to provide control of the slice energy spread~\cite{PhysRevSTAB.7.074401}. Downstream of the injector, the main accelerator consists of four accelerating sections and two bunch compressors, following a layout similar to that of the Linac Coherent Light Source II (LCLS-II)~\cite{osti_1029479} and the Shanghai High-repetition-rate XFEL and Extreme Light Facility (SHINE)~\cite{zhao2018shine}. The tunable range of the photon energy is limited by the adjustable range of the undulator magnetic field strength. Access to a broader photon-energy range therefore requires changing the electron beam energy, which would affect all downstream undulator lines. For simultaneous operation of multiple FEL lines, bunch-to-bunch electron-energy control is therefore needed, but remains challenging in a CW XFEL. To address this, a delay/compression section is placed before the final linac section L4 to support multienergy operation~\cite{PhysRevAccelBeams.22.090701,photonics12030275}. In the present design, this section consists of four horizontal double-bend achromats (DBAs). By changing the beam arrival time into L4, the electrons experience different off-crest accelerating phases, which lead to different output beam energies and allow simultaneous access to a broader photon-energy range in the downstream undulator lines. The same section can also provide additional compression with bunch-by-bunch control.

After the final linac, the beam is distributed to different FEL lines through switchyard sections. These switchyard sections should not be regarded merely as passive transport lines, but can also be designed as active compression sections with flexible control of the $R_{56}$ before the undulator. One possible switchyard design is shown in Fig.~\ref{Fig:part4_arc_design}. This design is based on two DBAs and additional micro-bends, allowing the transport section to provide flexible $R_{56}$ control from negative to positive while maintaining zero transverse dispersion downstream~\cite{PhysRevAccelBeams.27.110702,mnwd-ypfs,yan2025attoshine}. Such flexibility makes it possible to perform further bunch compression before the undulator and thereby generate a narrow high-current spike for attosecond lasing. Within this overall layout, three FEL lines may be designed to meet different user requirements and to provide flexibility for the future development of new attosecond modes.

The first FEL line is intended primarily for external seeding, while also supporting mode-locking operation. It begins with a sequence of modulation sections, each followed by a dispersive section, enabling seeded schemes such as HGHG~\cite{PhysRevA.44.5178} and EEHG~\cite{stupakov2009using}. External seeding can provide high temporal coherence and reproducible attosecond structures. HGHG has already been shown to support attosecond pulse-train generation~\cite{Maroju2020Nature}, while EEHG has been proposed for isolated attosecond pulse generation~\cite{Xiang2009PRSTAB,Yan2010NIMA,Xiao2022UFScience}. A complementary seeded route for attosecond pulse generation is to use strong single-stage energy modulation and radiate in a short radiator, namely a quasi-HGHG scheme~\cite{Deng2010CPC,Garcia2016PRAB}. This line can also serve as a platform for exploring few-cycle attosecond pulse generation~\cite{Tibai2014PRL}. With chicanes inserted between undulator modules, it can further operate in a mode-locking configuration for generating attosecond pulse trains~\cite{Thompson2008PRL,wn8d-l7sh}, and potentially even zeptosecond X-ray pulses~\cite{Dunning2013PRL}.

The second FEL line is intended for ESASE-based operation. Before the undulator, it contains a modulation section followed by a chicane to generate a narrow current spike. In addition, chicanes inserted between undulator sections can enable the generation of twin attosecond XFEL pulses with adjustable delay and photon energy, thereby supporting X-ray attosecond pump-probe experiments aimed at resolving ultrafast electronic processes in real time~\cite{Guo2024,li2024science}. Because ESASE enables relatively direct control of the current spike, this line is also well suited to multistage attosecond amplification~\cite{tanaka2016using,PhysRevLett.120.264801}. It can further be combined with orbital angular momentum control for spatiotemporal pulse shaping~\cite{xu2026spatiotemporalshapingattosecondxrays}. 

The third FEL line is intended for the highest peak power. It can operate in a self-chirping mode~\cite{yan2024terawattfel,yan2025attoshine} and with cathode-shaping schemes~\cite{Zhang2020NJP} for both high-charge and low-charge beams. Chicanes inserted between undulator sections can enable the generation of twin attosecond pulses with adjustable delay and can also support the multi-stage amplification mode. In this context, a larger energy spread in the second stage may help suppress the trailing tail and thereby enhance single-spike probability in the post-saturation superradiant regime~\cite{yang2020postsaturation}.

Beyond the accelerator, the full performance of an attosecond XFEL facility also depends on the photon transport system downstream of the undulators. The transport optics should provide high transmission while minimizing temporal stretching of the pulses. Nanofocusing optics, such as advanced Kirkpatrick–Baez mirrors~\cite{Yamada2024NatPhoton, Inoue2025Optica}, are important for maximizing the photon density at the sample.

\bibliography{ssoam}

\end{document}